\documentclass[]{aa}  
\usepackage{amsmath}
\usepackage{amssymb}
\usepackage{natbib}
\usepackage{graphicx}
\usepackage{txfonts}
\usepackage[english]{babel}
\usepackage{hyperref}
\usepackage{units}
\begin{document}
\title{The effects of disk and dust structure on observed polarimetric images of protoplanetary disks}


\author{
	M. Min\inst{1}
		\and
	H.~Canovas\inst{1}
		\and
	G.~D. Mulders\inst{2,3}
		\and
	C.~U. Keller\inst{1}
}

\offprints{M. Min, \email{M.Min@uu.nl}}

\institute{
Astronomical institute Utrecht, University of Utrecht, 
PO Box 80000, 3508 TA Utrecht, The Netherlands
	\and
Astronomical Institute ÒAnton PannekoekÓ, University of Amsterdam, 
PO Box 94249, 1090 GE Amsterdam, The Netherlands
	\and
SRON Netherlands Institute for Space Research, 
PO Box 800, 9700 AV, Groningen, The Netherlands
}

   \date{Last revision: \today}

 
  \abstract
   {Imaging polarimetry is a powerful tool for imaging faint circumstellar material. It is a rapidly developing field with great promise for diagnostics of both the large-scale structures and the small-scale details of the scattering particles.}
   {For a correct analysis of observations we need to fully understand the effects of dust particle parameters, as well as the effects of the telescope, atmospheric seeing, and assumptions about the data reduction and processing of the observed signal. Here we study the major effects of dust particle structure, size-dependent grain settling, and instrumental properties.}
   {We performed radiative transfer modeling using different dust particle models and disk structures. To study the influence of seeing and telescope diffraction we ran the models 
   through an instrument simulator for the ExPo dual-beam imaging polarimeter mounted at the 4.2m William Herschel Telescope (WHT).}
   {Particle shape and size have a strong influence on the brightness and detectability of the disks. In the simulated observations, the central resolution element also contains contributions from the inner regions of the protoplanetary disk besides the unpolarized central star. This causes the central resolution element to be polarized, making simple corrections for instrumental polarization difficult. This effect strongly depends on the spatial resolution, so adaptive optics systems are needed for proper polarization calibration.}
   {We find that the commonly employed homogeneous sphere model gives results that differ significantly from more realistic models. For a proper analysis of the wealth of data available now or in the near future, one must properly take the effects of particle types and disk structure into account. 
   The observed signal depends strongly on the properties of these more realistic models, thus providing a potentially powerful diagnostic. We conclude that it is important to correctly understand telescope depolarization
   and calibration effects for a correct interpretation of the degree of polarization.}


   \maketitle
%

\section{Introduction}

High spatial-resolution imaging polarimetry is a rapidly developing field. It is now recognized that imaging polarimetry is a powerful tool for imaging faint matter around a bright star. Polarimetric differential imaging (PDI) is used to suppress the glare of the central star, which is unpolarized, to be able to see the faint reflected light from the material surrounding the star, which becomes polarized by scattering. This has already been successfully used to image circumstellar disks around young stars \citep[see e.g.][]{1997ApJ...489..210C, 2011ApJ...729L..17H} and circumstellar dust shells around evolved stars \citep[e.g][]{2001MNRAS.322..321G}, and it has been proposed for imaging the light reflected from exoplanets \citep{2004A&A...428..663S}. Images in polarized intensity not only contain information on the geometry of the circumstellar material, but also carry information on the particles scattering the light from the central star. In the general case, these two characteristics are intertwined and not easily separated.

Perhaps the most challenging case for interpreting polarimetric images is that of optically thick circumstellar disks around newly born stars. First, being optically thick causes multiple scattering, which alters the degree of polarization. The geometry of the surface of these disks can also cause scattering at unexpected angles, in turn causing certain features in polarization \citep{2009ApJ...707L.132P}. These effects can complicate the interpretation of polarimetric images in terms of planet-disk interactions. Besides these geometrical effects on a large scale, a second complicating factor arises when these images are used to derive general information about the dust particles in the disk. Dust particles are expected to be highly complex in shape and might be quite large in size, since they are a mixture of small grains and large, complex fluffy aggregates \citep{2009ASPC..414..494D}. Computing the optical properties of such particles is very difficult and computationally demanding, if feasible at all.  In addition, the different grain sizes are not expected to be mixed well with the gas, but size-dependent settling of the dust particles causes a size distribution of the grains that changes with height in the disk \citep{2004A&A...421.1075D}. To conclude this chain of complications, dust properties, i.e. grain size and composition, are expected to vary radially \citep[see e.g.][]{2004Natur.432..479V}.
Previous studies have already covered large parts of the parameter space of dusty circumstellar disks in the context of polarimetric imaging. The effects of disk mass and radial extend were studied in e.g. \citet{1992ApJ...395..529W}. More recently, \citet{2010A&A...518A..63M} has shown computations of near infrared polarimetric imaging of disks with various grain sizes and parameterized thickness of the disk.

An additional complication is the effect of the finite resolution of real telescopes on the final image. The most powerful diagnostic, the degree of polarization of the scattered light, is highly influenced by the glare from the central star. Also, blurring of the high-intensity, polarized, inner regions of the disk will affect the appearance of the disk in polarized intensity. Furthermore, data reduction techniques that correct for instrumental polarization using the central star as an unpolarized calibration point might suffer from the non-zero innermost disk polarization.

In this paper we present computations of the polarimetric images of protoplanetary disks around young, intermediate-mass stars, so-called Herbig Ae stars. We study how the images are effected by the structure of the dust grains and size-dependent grain settling. We varied the size of the dust particles, the geometry of the disk, and the shape of the dust particles. This was done using state-of-the-art models for the optical properties of the dust particles combined with a highly efficient radiative transfer code. An instrument simulator for a dual-beam imaging polarimeter at a 4\,m class telescope was used to study the effects of atmospheric seeing and data reduction artifacts.

\section{Model setup and computational approach}
\label{sec:computational approach}

In this section we describe the setup of the models in terms of dust grains and disk geometry. {in Section \ref{sec:par space} we motivate our choice for the parameters used in the computations.} The radiative transfer model used to create the polarimetric images and the instrument simulator are also described.

\subsection{Parameter space}
\label{sec:par space}

{The characteristics of dusty, protoplanetary disks are described by a large set of parameters ranging from the microscopic properties of the dust grains to the macroscopic geometrical properties of the disk. 
\citet{1992ApJ...395..529W} investigated the effects of disk mass, radial extent, and the amount of disk flaring. They conclude that a flat disk would not be detectable. Also, they conclude that the large extent of many observations available at that time was most likely caused by an extended infalling envelope. The appearance of such an envelope in combination with the effects of outflows was studied in \citet{1993ApJ...402..605W} and \citet{1996A&A...308..863F} for a wide range of envelope and outflow parameters. 
More recently, \citet{2010A&A...518A..63M} has presented computations of near infrared polarimetric imaging of disks. 
These theoretical studies all provide a good analysis of the dependence of polarimetric imaging on certain parameters of the dusty surroundings of young stars. It is therefore not the aim of this paper to repeat this and present a study covering the entire parameter space. 
}

{There are a few aspects that have not received much attention in previous studies, so we focus on these aspects and study their effects on the observed polarimetric images. These aspects are
\begin{itemize}
\item Grain structure,
\item Size dependent grain settling,
\item Atmospheric blurring, and other observational aspects.
\end{itemize}
Below we explain how we set the model up specifically keeping these three main points in mind.}

\subsection{Dust grain model}

\subsubsection{Composition}

The composition of protoplanetary dust is far from trivial \citep[for a recent review see][]{2010pdac.book..161M}. To arrive at a simple, yet quite accurate model of the composition, we follow the procedure also adopted by \citet{2011Icar..212..416M}. First we define the abundances of all available dust-forming elements to be the same as the solar composition derived by \citet{1998SSRv...85..161G}. In meteorites and interstellar dust particles (IDPs) in our solar system all of the available sulfur is in the form of iron sulfide. Therefore, we first put all S in FeS thereby taking away about half of the available Fe atoms. From the remaining Fe, Mg, and Si we produce an amorphous Fe/Mg silicate with a stoichiometry in between olivine and pyroxene. We model this mixture by averaging the optical properties of different types of amorphous silicates. In addition to this iron sulfide/silicate mixture we add a contribution from carbonaceous matter. It is found that in our solar system about half of the carbon atoms are in the solid phase
\citep{1987A&A...187..859G}, so we add this abundance in the form of amorphous carbon grains. The composition resulting from this contains by mass 58\% silicates, 18\% iron sulfide, and 24\% amorphous carbon.

To compute the opacities of the dust grains, we need the refractive index as a function of wavelength. This has been measured in the laboratory. For the silicates we use the measurements by \citet{1995A&A...300..503D} and \citet{1996A&A...311..291H}, for the iron sulfide we take the laboratory data from \citet{1994ApJ...423L..71B}, and for the carboneceous material we take the measurements from \cite{1993A&A...279..577P}.

{It can be argued that grains in the outer regions of protoplanetary disks contain a coating of ice. However, there is a recent debate about whether ice layers can survive in the upper layers of the disk where it is exposed to UV radiation from the central star \citep[see][]{2007A&A...475..755G}. We checked how the scattering properties are influenced when adding water ice to the mixture in an abundance of 45\% by mass, which is the abundance expected when putting all remaining oxygen in water ice \citep{2011Icar..212..416M}. The scattering properties are affected only slightly. The single scattering albedo goes up by less than 15\% in all cases. The extinction coefficient, the maximum degree of polarization, and the asymmetry of the phase function are hardly affected. The resulting polarimetric images are thus very similar to those without ice.}

\begin{figure*}[!htb]
\begin{center}
\resizebox{0.8\hsize}{!}{\begin{tabular}{ccc}
\includegraphics{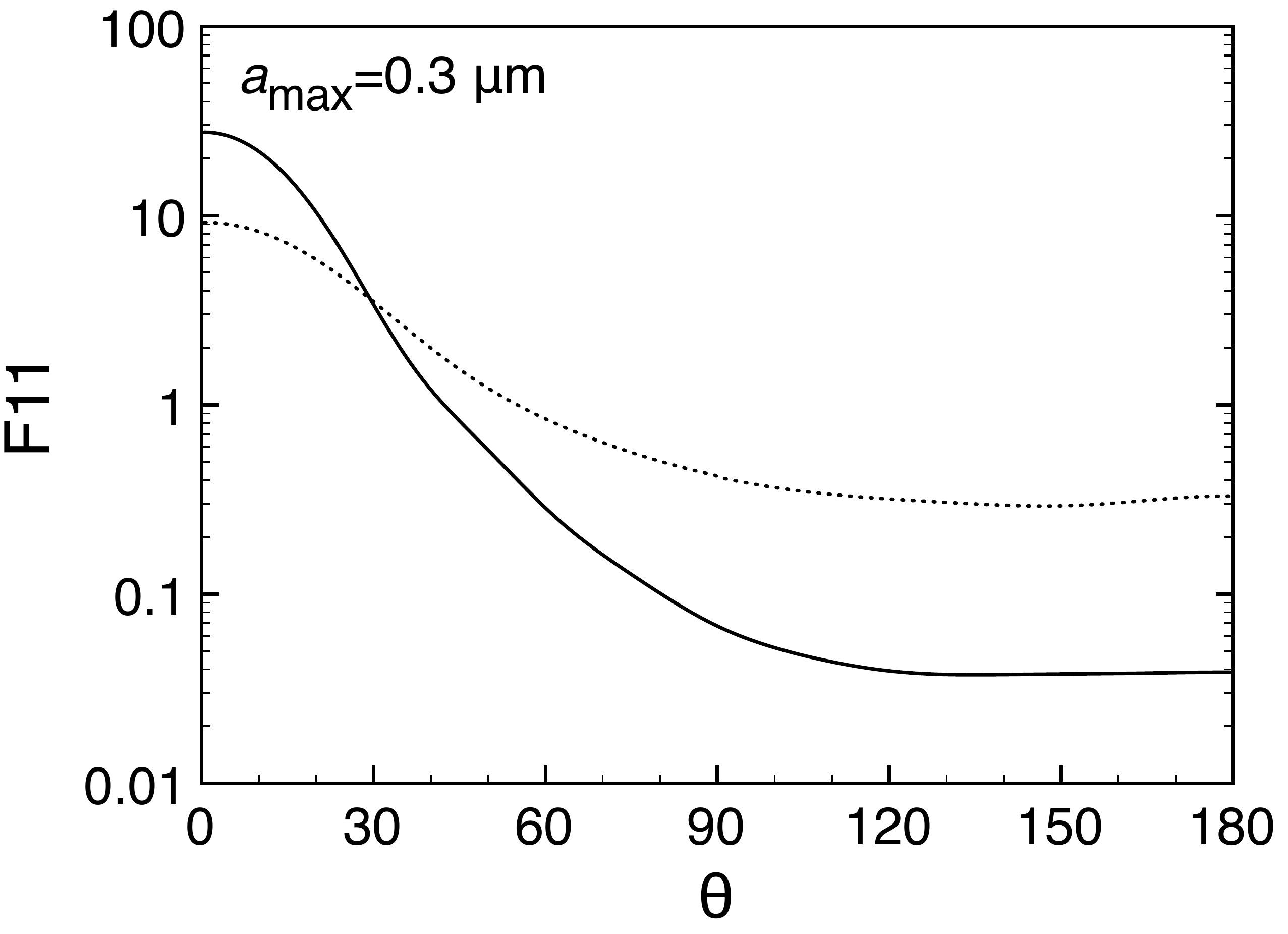} &
\includegraphics{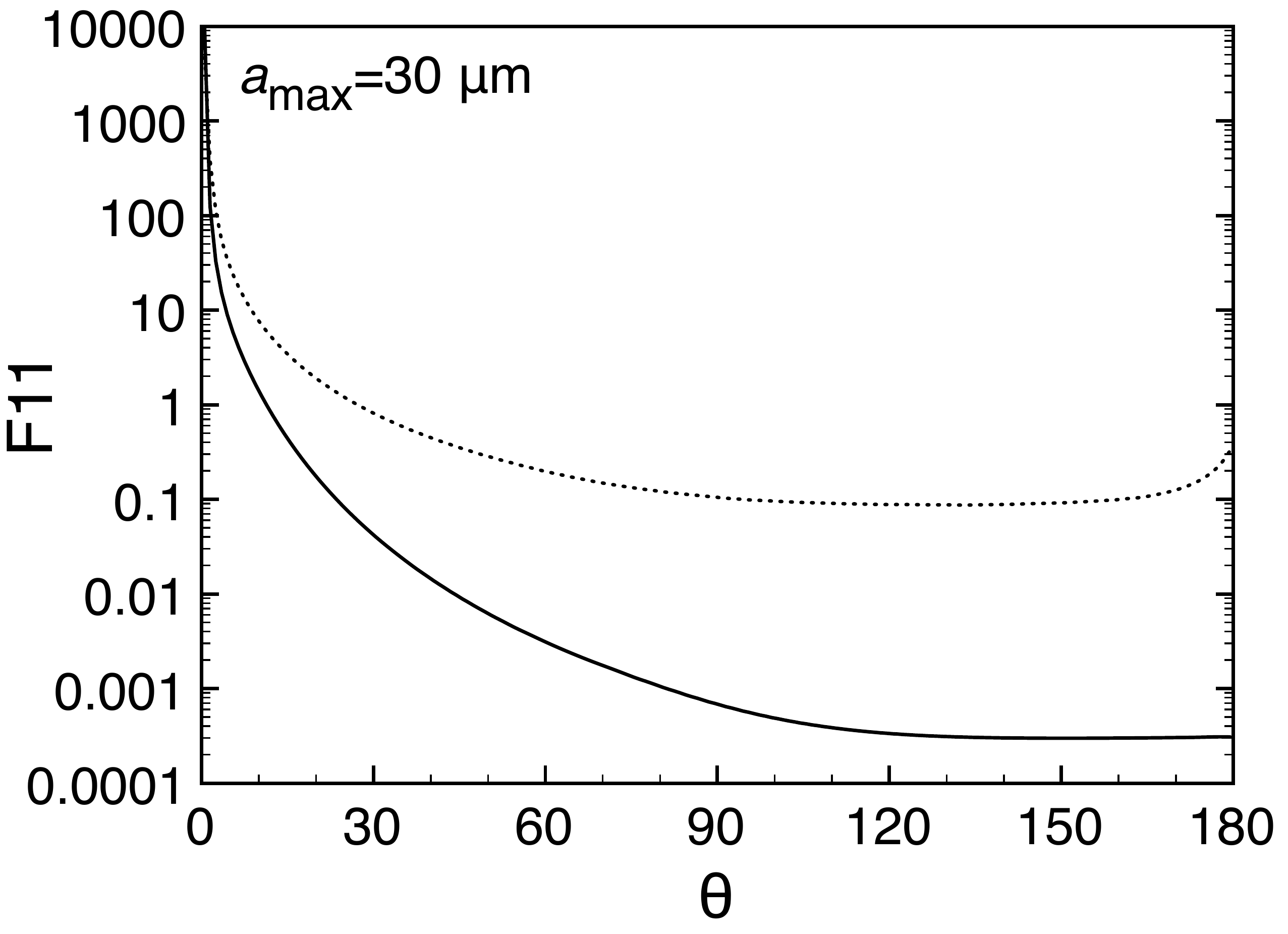} &
\includegraphics{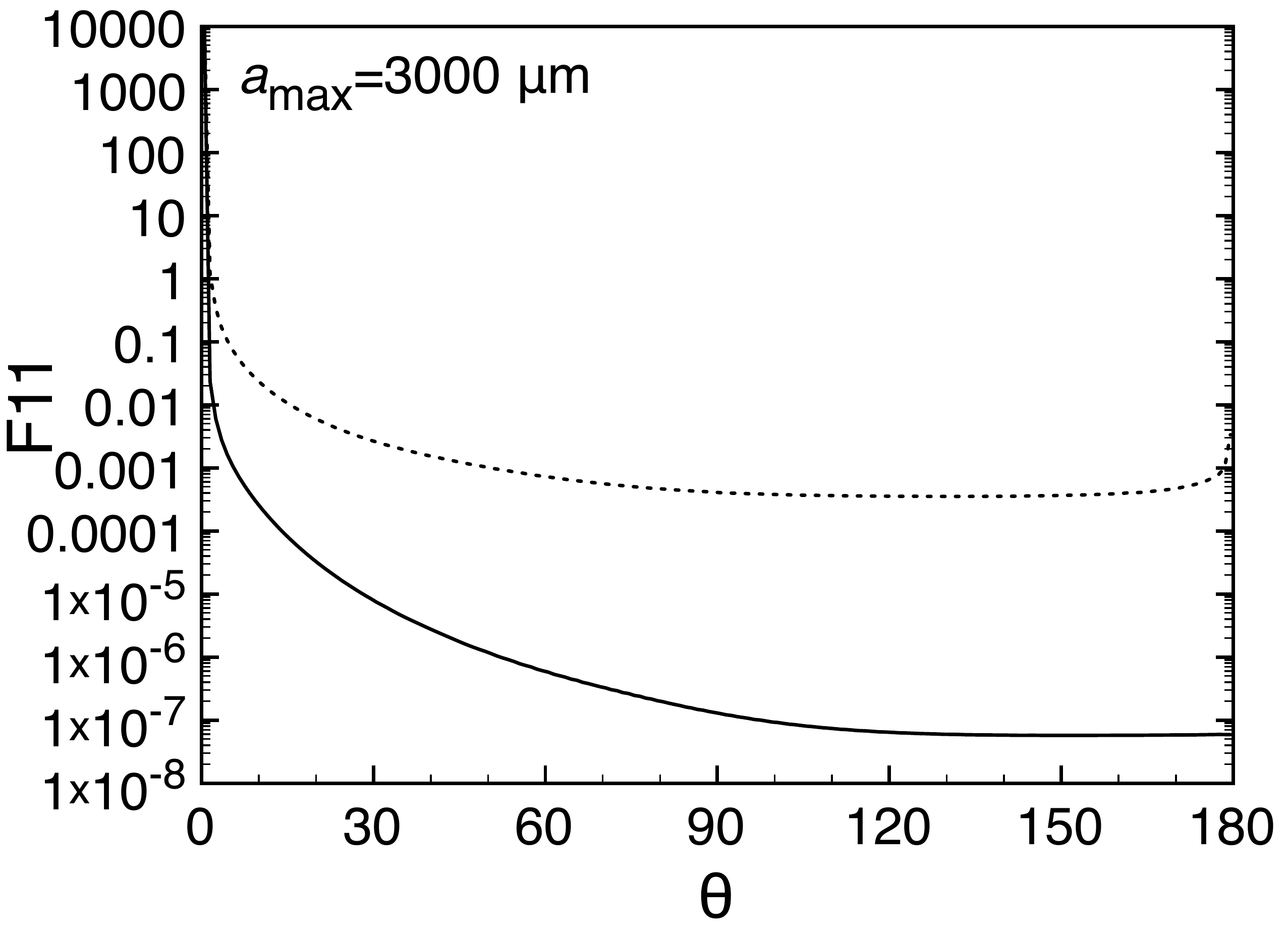} \\
\includegraphics{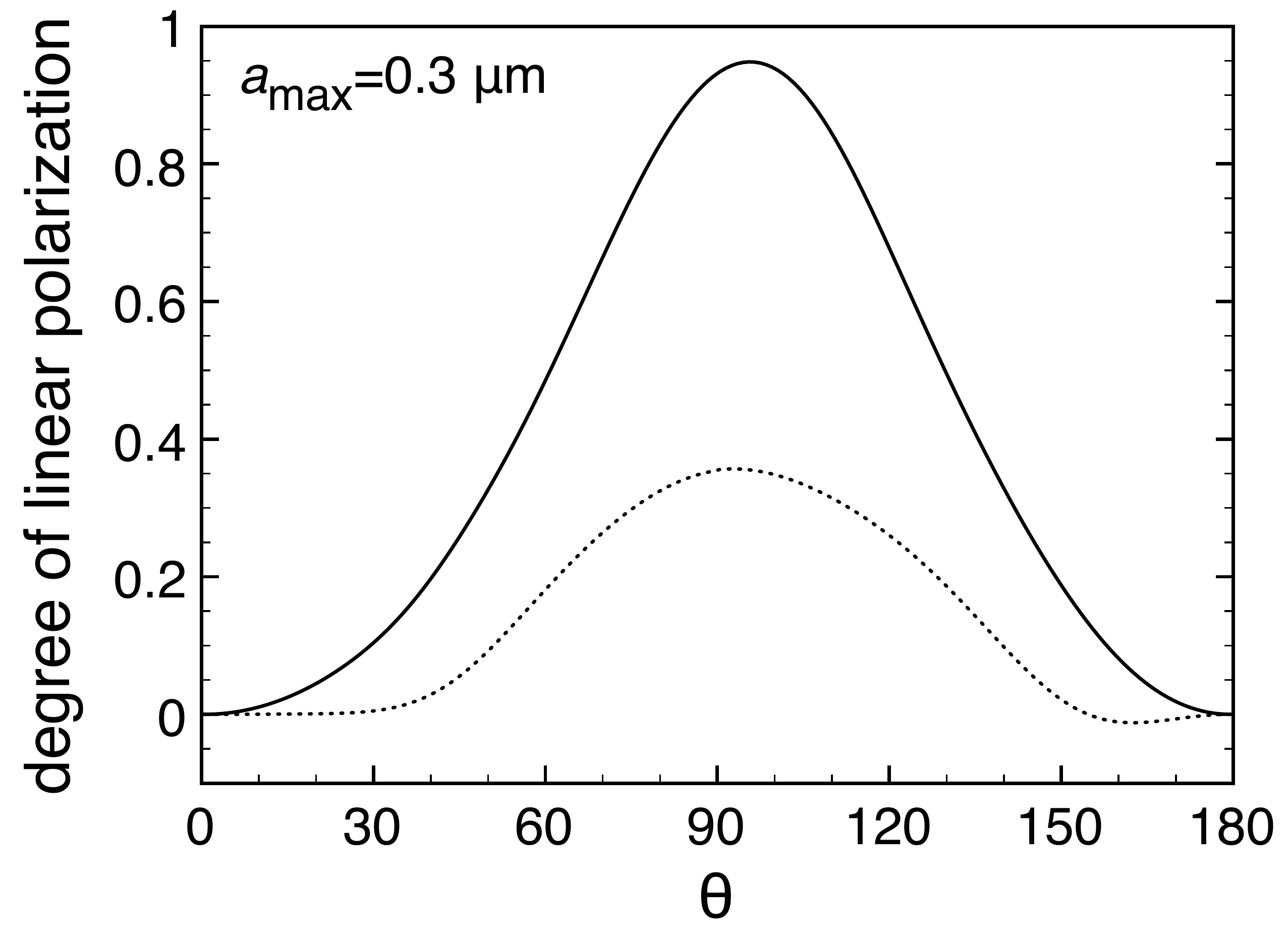} &
\includegraphics{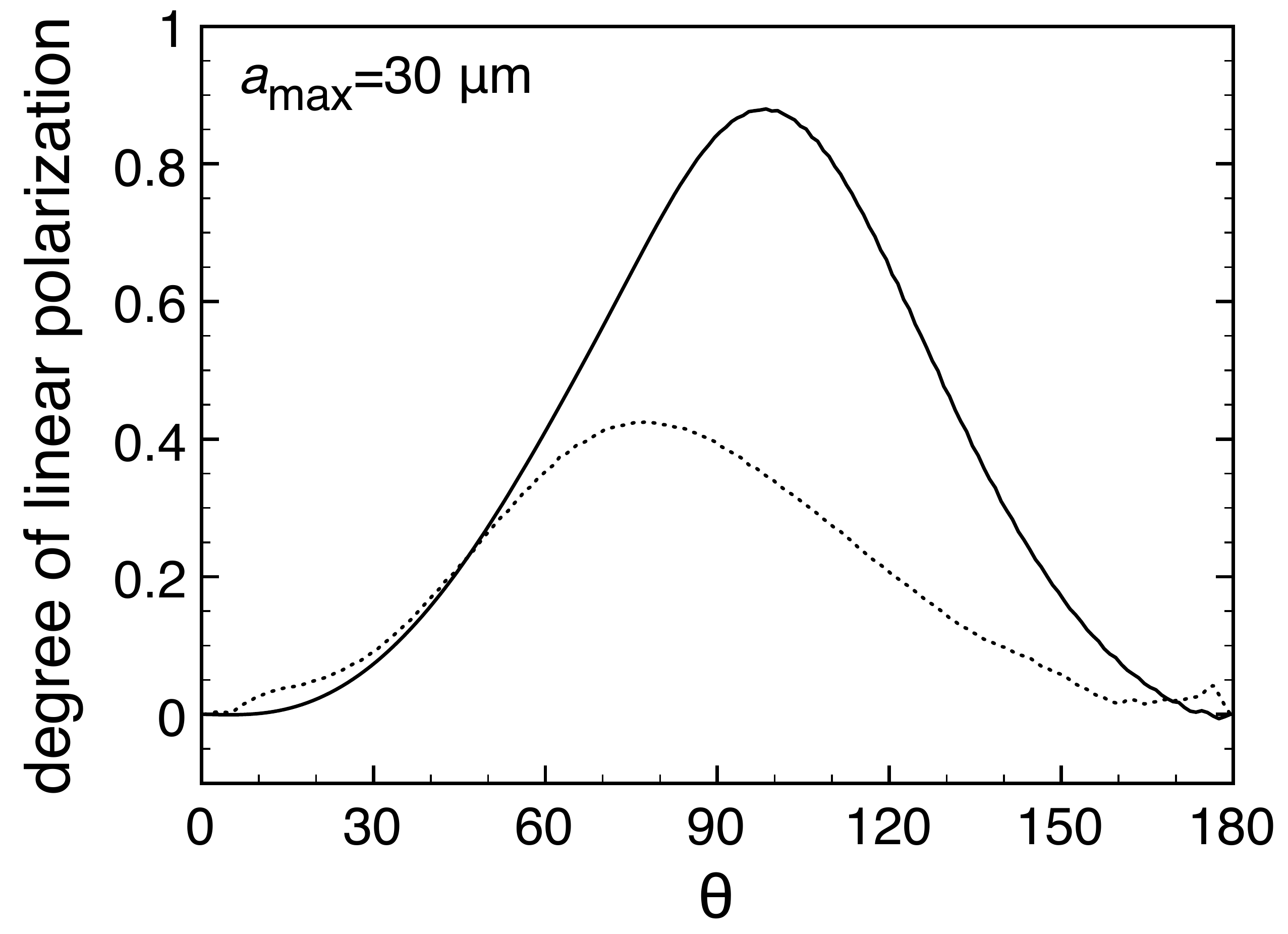} &
\includegraphics{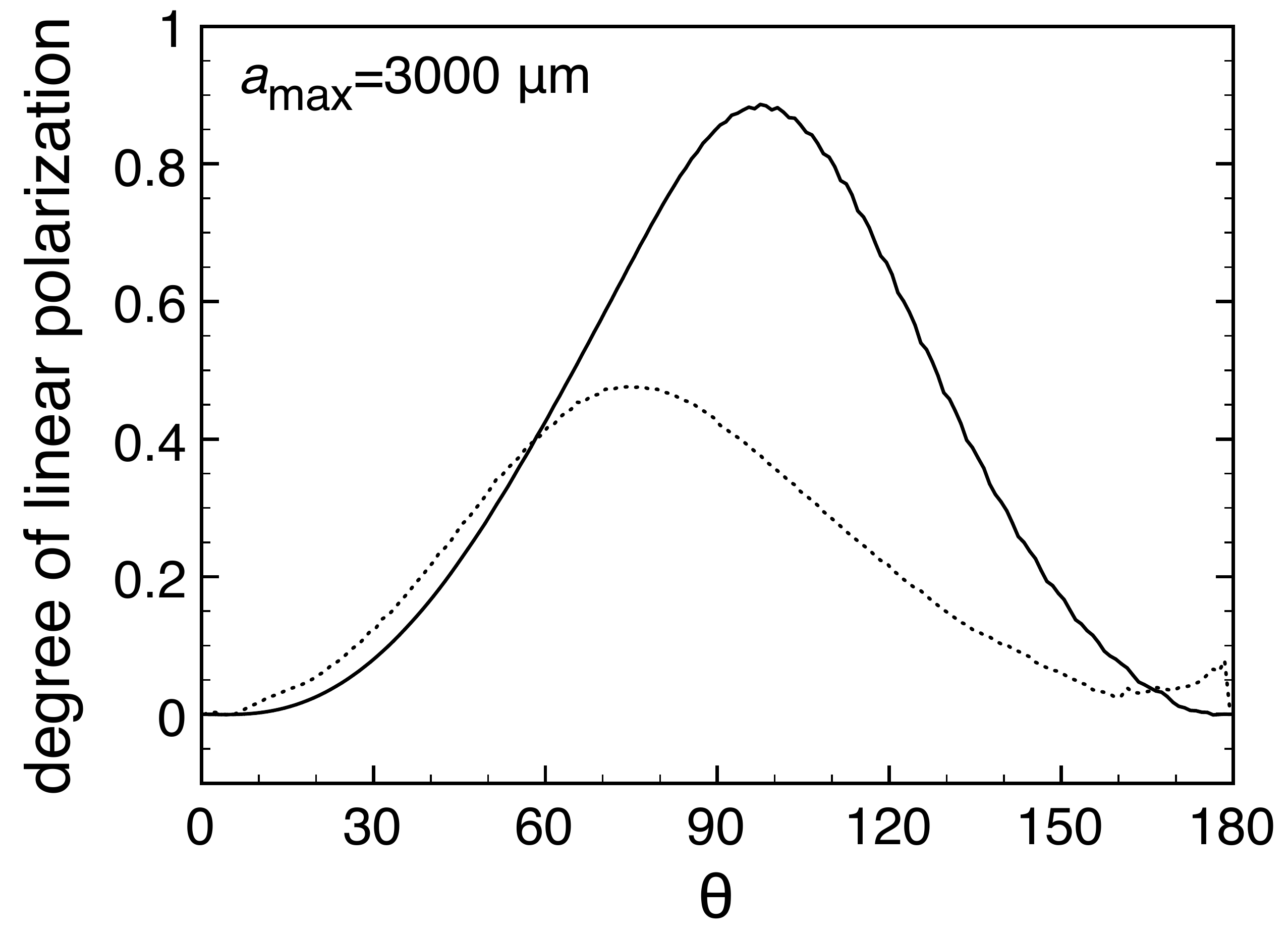} \\
\end{tabular}}\end{center}
\caption{The scattering properties of the dust particles for different values of the maximum particle size, $a_\mathrm{max}$. Upper panels: the normalized phase function, $F_{11}$, as a function of scattering angle, $\theta$. Lower panels: the degree of linear polarization as a function of scattering angle. The wavelength is $\lambda=600\,$nm. Solid lines are for the fluffy aggregates, dashed lines for the compact grains.}
\label{fig:matrix}
\end{figure*}

\subsubsection{Grain structure}

We distinguish between two extreme types of particle structures using the above composition:
\begin{enumerate}
\item compact, homogeneous grains, and
\item fluffy, inhomogeneous aggregates.
\end{enumerate}
These two different particle structures require different ways of computing the optical properties. To correctly estimate the scattered light intensity and degree of polarization, the model used for the optical properties of the grains is crucial. Homogeneous spherical grains cause resonances at certain scattering angles and generally behave differently from more natural, irregular shapes. However, for reference and comparison to previous work, we also computed several models using compact, homogeneous spheres.

\emph{Compact, homogeneous grains:} This model considers a mixture of grains, each composed of a single material. The optical properties of such a mixture can be obtained by simply averaging the optical properties of homogeneous grains using the given abundances of the various materials. We chose here to model the optical properties of the dust grains using the statistical approach. In this approach we take a wide distribution of simple particle shapes to simulate the variety of irregularities occurring in nature. We use the distribution of hollow spheres \citep[DHS; see][]{2005A&A...432..909M} with an `irregularity parameter' $f_\mathrm{max}=0.8$. It turns out that this choice of this grain-shape model, together with our choices for the grain sizes and compositions, provides a scattering matrix that is quite similar to cometary dust.

\emph{Fluffy, inhomogeneous aggregates:} When small dust grains collide, they can form large aggregated structures. The optical properties of such aggregates display characteristics of both the constituents and the size of the aggregate as a whole \citep[see e.g.][]{2007A&A...470..377V}. We use here the approximate method by \citet{2008A&A...489..135M} to compute the optical properties of large, inhomogeneous fluffy aggregates. In this method we use effective medium theory to mix the dust material with vacuum and thus obtain a highly porous, fluffy structure. The vacuum fraction can be computed from the aggregate size and fractal dimension. Though this method was not primarily designed to simulate scattering at optical wavelengths, we find that many of the characteristics attributed to fluffy aggregates are simulated well. In general we find that the grains computed using these methods with the parameters given by \citet{2008A&A...489..135M} are extremely fluffy and as such give an extremely forward-peaked scattering phase function and a very high degree of polarization at a 90$^\circ$ scattering angle.

\subsubsection{Grain size}

To address the effects of grain growth on the observable polarimetric signal we use a parameterized grain size distribution. We adopt a power law size distribution with $n(a)\propto a^{-3}$. Here $a$ is the volume equivalent radius of the grains and the distribution runs from $a_\mathrm{min}$ to $a_\mathrm{max}$. We fix $a_\mathrm{min}$ to $0.03\,\mu$m and vary $a_\mathrm{max}$ from $0.3\,\mu$m to $3\,$mm. The different grain sizes are treated separately in the radiative transfer. This implies that they can have different temperatures and a different vertical distribution caused by grain settling.

Figure~\ref{fig:matrix} displays the single scattering phase function and the degree of linear polarization as functions of the scattering angle, $\theta$, for different values of $a_\mathrm{max}$ at a wavelength of $\lambda=0.6\,\mu$m. The most important scattering characteristics are summarized in Table~\ref{tab:scatt parameters}. Shown are the mass extinction coefficient, $\kappa_\mathrm{ext}$, the single scattering albedo, $\omega$, the asymmetry parameter of the phase function, $\left<\cos\theta\right>$, and the maximum degree of polarization from single scattering, $P_\mathrm{max}$.

\begin{table}[!tbp]
\begin{center}
\begin{tabular}{lccccc}
Particle type & $a_\mathrm{max}$	& $\kappa_\mathrm{ext}$	& $\omega$	& $\left<\cos\theta\right>$	& $P_\mathrm{max}$ \\
			& [$\mu$m]					&	[cm$^2$/g]		&			&					& [\%] \\
\hline
Compact grains		& $0.3$	&	35290	&	0.63	&	0.53		&	33	\\
				& $30$	&	1532		&	0.73	&	0.76		&	40	\\
				& $3000$	&	25		&	0.71	&	0.79		&	44	\\
Aggregates		& $0.3$	&	31246	&	0.51	&	0.87		&	95	\\
				& $30$	&	5277		&	0.57	&	0.98		&	89	\\
				& $3000$	&	139		&	0.53	&	0.99		&	89	\\
\end{tabular}
\end{center}
\caption{Light scattering characteristics of the different particle types for a few values of $a_\mathrm{max}$.}
\label{tab:scatt parameters}
\end{table}

\subsection{Disk model}

For the disk model we try to avoid parameterization as much as possible. Therefore we compute the vertical scale height of the disk self-consistently. {This means that the vertical density distribution is fully determined by the temperature structure of the disk assuming vertical hydrostatic equilibrium (see \citealt{2007prpl.conf..555D} for a review of the theoretical picture of the structures of protoplanetary disks).} In addition, we compute the radial location and structure of the inner rim using evaporation and sublimation physics as described in \citet{2009A&A...506.1199K}. {This results in an inner edge of the disk that is curved outwards and slightly higher than the regions behind it. For details we refer to \citet{2009A&A...506.1199K}.}

\begin{table}[!tbp]
\begin{center}
\begin{tabular}{llc}
parameter				&	symbol		&	value \\
\hline
mass of the star		&	$M_\star$			&	$2.5\,M_{\sun}$ \\
effective temperature	&	$T_\mathrm{eff}$	&	$10000\,$K \\
radius of the star		&	$R_\star$			&	$2\,R_{\sun}$ \\
dust mass in the disk		&	$M_\mathrm{dust}$	&	$1\cdot10^{-3}\,M_{\sun}$ \\
distance to the star		&	$D$				&	$150\,$parsec \\
outer radius of the disk	&	$R_\mathrm{out}$	&	$2000\,$AU \\
surface density turnover point	&	$R_0$		&	$200\,$AU \\
minimum grain size		&	$a_\mathrm{min}$	&	$0.03\,\mu$m\\
maximum grain size		&	$a_\mathrm{max}$	&	$0.3, 30, 3000\,\mu$m\\
turbulence parameter	&	$\alpha$			&	$10^{-4}$ (settled), $\infty$ (flared)\\
gas-to-dust ratio		&	$f_{g/d}$			&	$120$\\
\end{tabular}
\end{center}
\caption{Values for the model parameters}
\label{tab:parameters}
\end{table}

\begin{figure*}[!htb]
\centerline{\resizebox{0.9\hsize}{!}{\includegraphics{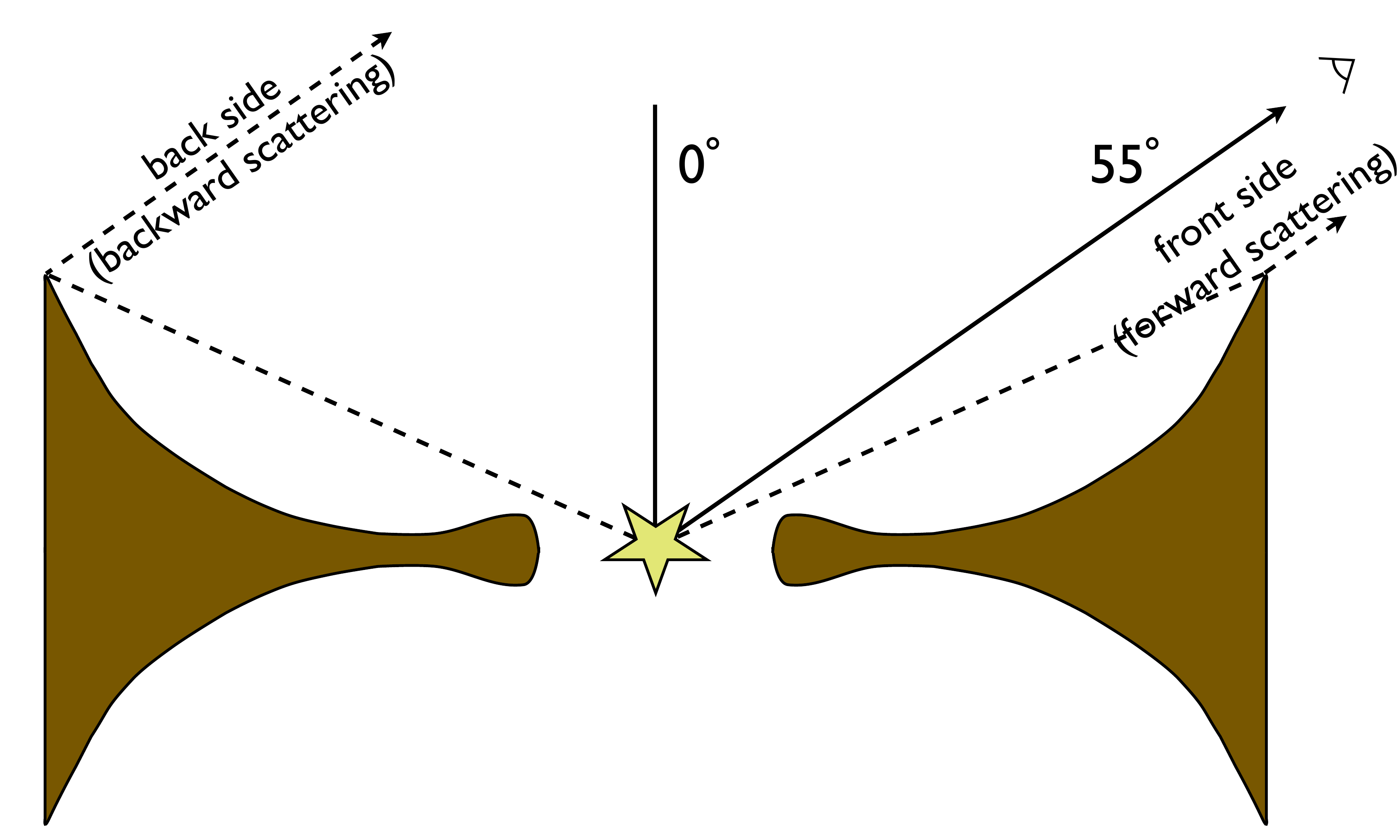}\includegraphics{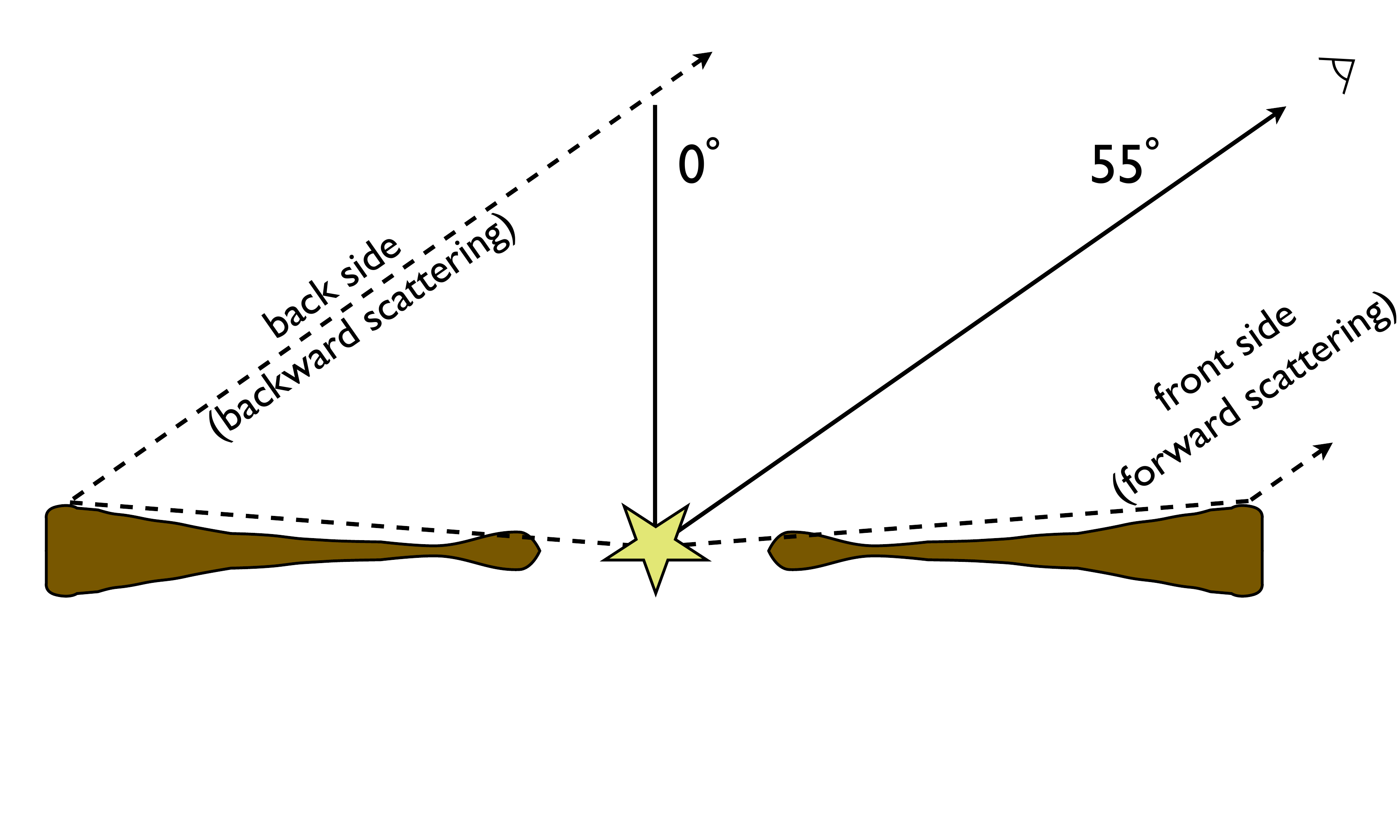}}}
\caption{Sketch of the two disk geometries. The left panel shows the flaring, well-mixed disk structure, while the right panel depicts the settled, flat disk.}
\label{fig:disk cartoon}
\end{figure*}

For the central star we take a Herbig Ae star. We use two different disk types, a flaring disk, where we assume gas and dust to be perfectly coupled, and a settled disk, where we use the settling prescription from \citet{1995Icar..114..237D} as described above. These two disk types approximately represent the Group I and II sources as defined by \citet{2001A&A...365..476M}. All star and disk parameters are listed in Table~\ref{tab:parameters}.

As the grains grow to larger sizes they decouple from the gas and settle towards the midplane \citep{2004A&A...421.1075D}. This is thought to be a possible explanation for the flatter disk structures observed around some stars. A full treatment of settling is beyond the scope of this paper and will be implemented in the future in the MCMax radiative transfer code. Here we parameterize settling using the approximate method by \citet{1995Icar..114..237D}. In this way we do simulate the dependence of settling efficiency on grain size, grain density, gas density, and gas temperature. To distinguish between effects of grain settling and grain optical properties we take for the grain settling parameters always those of the material volume equivalent sphere. In principle, settling is more efficient for homogeneous compact grains and less efficient for fluffy aggregates. This is simply because the larger surface area of the fluffy grains allows them to couple much more easily to the gas. However, since we want to be able to separate disk structure effects as much as possible from effects of grain properties, we have chosen to not let this effect influence the settling efficiency and determine the settling properties of all grains as if they were spherical homogeneous grains with the same mass. The description of grain settling we use has one free parameter, which is the strength of the turbulence, $\alpha$. This parameter defines the balance between settling and vertical mixing. For the settled disk models we take $\alpha=10^{-4}$, while for the well mixed models we set $\alpha=\infty$.

The radial density distribution of the dust disk was parameterized using a radial surface density \citep{2008ApJ...678.1119H}
\begin{equation}
\Sigma(r) \propto r^{-p}\exp \left\{- \left( \frac{r}{R_0} \right) ^{2-p}\right\},
\end{equation}
for $r < R_\mathrm{out}$. Here $R_0$ is the turnover
point from where an exponential decay of the surface density sets in
and $p$ sets the power law in the inner region. We fix this  to
$p=1$, which is a commonly used value \citep[see e.g.][]{2006ApJ...640L..67D}. Although we take the outer radius of the disk to be 2000\,AU, the surface density of the disk is already highly diminished after the turnover radius $R_0=200\,$AU.

In Fig.~\ref{fig:disk cartoon} we depict the two different disk structures considered. This figure illustrates that the scattering angles of the surface layers of the disk depend strongly on the opening angle. In the case of a very flat disk, the scattering angles observed at the front and the back of the disk are symmetric around $90^\circ$. With $i$ the inclination angle of the disk the scattering angles at front and back for a very flat disk are $\theta=90^\circ-i$ and $\theta=90^\circ+i$ respectively. For a strong flaring disk with a large opening angle $\psi$, the scattering angles at front and back are symmetric around $90^\circ-\psi$ and become $\theta=90^\circ-\psi-i$ and $\theta=90^\circ-\psi+i$.

{We would like to emphasize that by using the self-consistent treatment described above we limit the amount of free parameters significantly. Besides the parameters we do vary, we have two important parameters that determine the observed polarization signal; $R_0$ and $M_\mathrm{dust}$. The impact of these two parameters on the polarimetric signal of circumstellar disks was studied in, e.g., \citet{1992ApJ...395..529W}.}

\subsection{Radiative transfer}

The transfer of radiation through the disk is done in full using the 3D axisymmetrical radiative transfer code MCMax \citep{2009A&A...497..155M}. This code uses Monte Carlo radiative transfer to compute the disk temperature structure and from this the vertical density scale-height and the sublimation state of the dust \citep[see also][]{2009A&A...506.1199K}. The scattering of radiation uses the full scattering matrix acting on the Stokes vectors. The grains in the disk are assumed to be in random orientation so symmetry arguments can be used to simplify the scattering matrix.

Often images and other observables are extracted from Monte Carlo radiative transfer methods by simply collecting the escaping photons after they are traced through the disk. Although this is an elegantly simple way, we adopt a different technique to construct the images which allows us to obtain accurate results even in the very faint parts of the image.
After the temperature structure of the disk is obtained, images are constructed by integrating the formal solution to the radiative transfer. For this, the local scattering source function is also needed; i.e., at each volume element in the disk we need to know how much radiation is scattered towards the observer. This is again computed using a Monte Carlo approach. However, this time we store the full 3D radiation field at each location in the disk by using the full path traveled by the Monte Carlo photons \citep[as inspired by the path length method for temperature determination by][]{1999A&A...344..282L}. In combination with the local thermal source function, which is obtained from the temperature, the formal solution can be easily integrated to obtain smooth images. This is done by integrating from each volume element in the disk the total emitted flux (as obtained from the local temperature) and the total scattered flux (as obtained from the 3D radiation field) diminished by the extinction through the disk along the line of sight towards the observer.

{The advantage of the above method over `classical Monte Carlo' photon-collecting methods is that the information accumulated during the Monte Carlo photon-tracing process is used to the fullest to obtain the local 3D radiation field. This way smooth images can also be produced in regions where only a small fraction of the photons arrive. This is especially important for imaging polarimetry where one of the main purposes is to gain a very high dynamic range in the image. The method was benchmarked against other radiative transfer codes in \citet{2009A&A...498..967P}.}

\begin{figure*}[!htb]
\centerline{\resizebox{0.9\hsize}{!}{\includegraphics{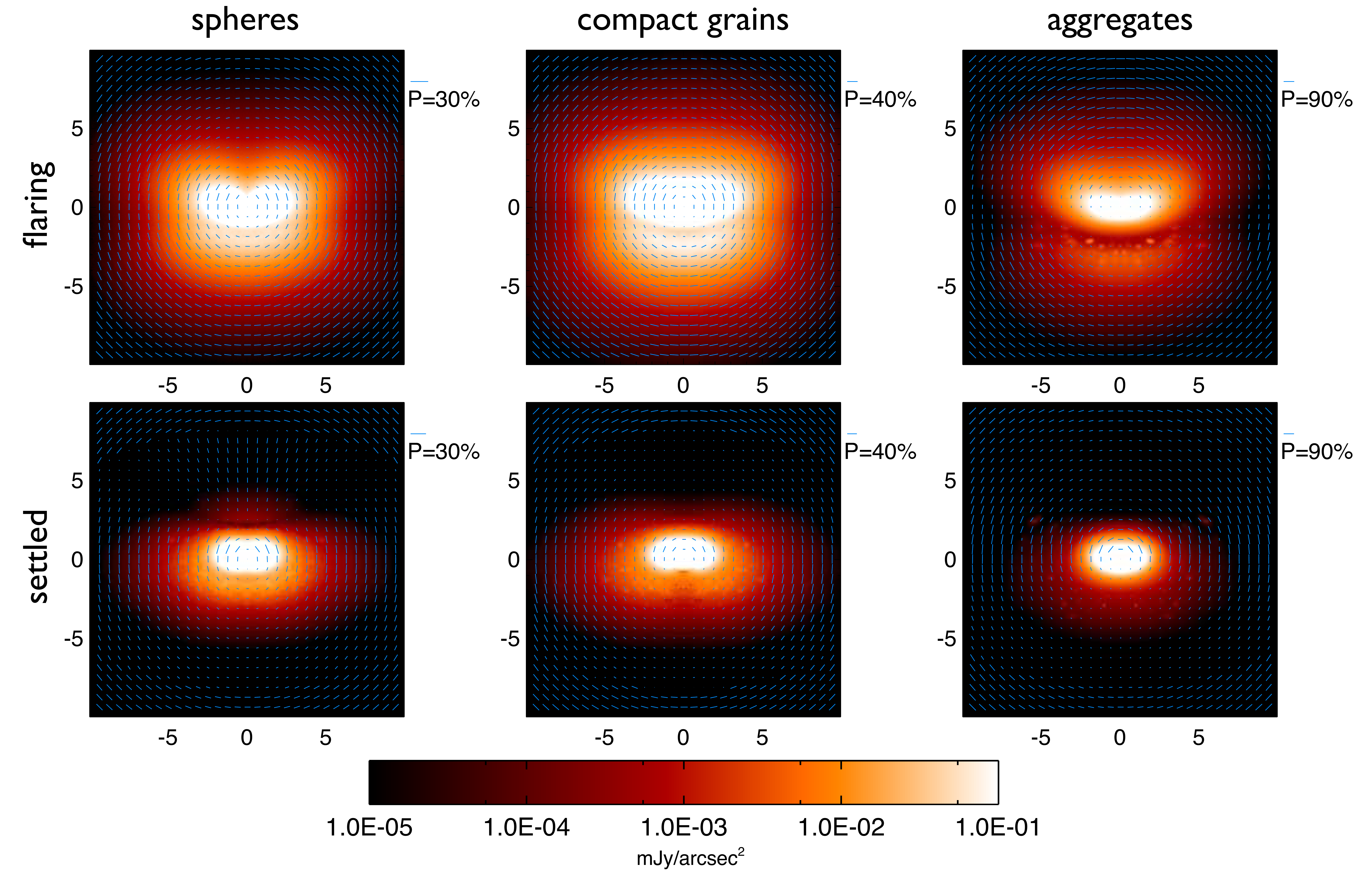}}}
\caption{Comparison of the polarized intensity maps for different grain types. Upper panels are for the flared disks, lower panels for the settled disks. Left panel: homogeneous spheres; middle panels: compact, irregularly shaped grains; right panels: fluffy aggregates. For all images, the maximum size of the particles is 30\,$\mu$m. Spatial scale is in arcseconds. The length of the vectors is proportional to the degree of polarization. The inclination of the disk is 55$^\circ$ with respect to pole on.}
\label{fig:grain types}
\end{figure*}

\subsection{Instrument simulation}

Ground-based observations are affected by atmospheric and instrumental effects such as seeing and instrumental polarization.
To account for these effects, a numerical simulation including readout noise, photon noise, instrumental polarization, and different
seeing conditions was performed. These simulations model the dual-beam imaging 
polarimeter ExPo \citep[the Extreme Polarimeter;][]{Rodens_2008,Rodens_2011}, currently a visitor instrument at the 4.2 meters William Herschel Telescope (WHT). 
ExPo measures linear polarization at optical wavelengths, and combines very short exposure times with the dual-beam technique to minimize systematic errors. In a standard observation, thousands of single exposure frames are combined to obtain a total exposure time of $\sim15$ minutes.

The point spread function (PSF) for ExPo is calculated as
\begin{equation}
\mathrm{PSF}(\lambda) = \left|  \mathcal{F} \{  A \cdot e^{i\varphi(\lambda)}   \}   \right|^{2},
\label{eq:psf}
\end{equation}
where $\mathcal{F}$ stands for \textit{fourier transform}, $A$ and $\phi$ refer to the amplitude and phase of the incoming wavefronts, respectively.
A set of 100 statistically independent wavefronts is calculated for different seeing conditions. 
We compute the phase of these wavefronts according to the Kolmogorov theory of turbulence (for details see Appendix \ref{sec:seeing}). Broadband PSFs are obtained from these sets of wavefronts combined with the aperture of the telescope after which they are convolved with full-resolution simulated images. Readout noise, photon noise, instrumental polarization, and guiding errors are added to these images.  This set of simulated-raw images is then reduced with the \textit{ExPo Pipeline}, described in detail in \citet{2011A&A...531A.102C}. 

The characteristics of this simulation are summarized in Table~\ref{tab:simterms}. The simulation of the atmosphere, telescope, and instrument is detailed in Appendix~\ref{app:instrument}.

\begin{figure*}[!htb]
\centerline{\resizebox{0.9\hsize}{!}{\includegraphics{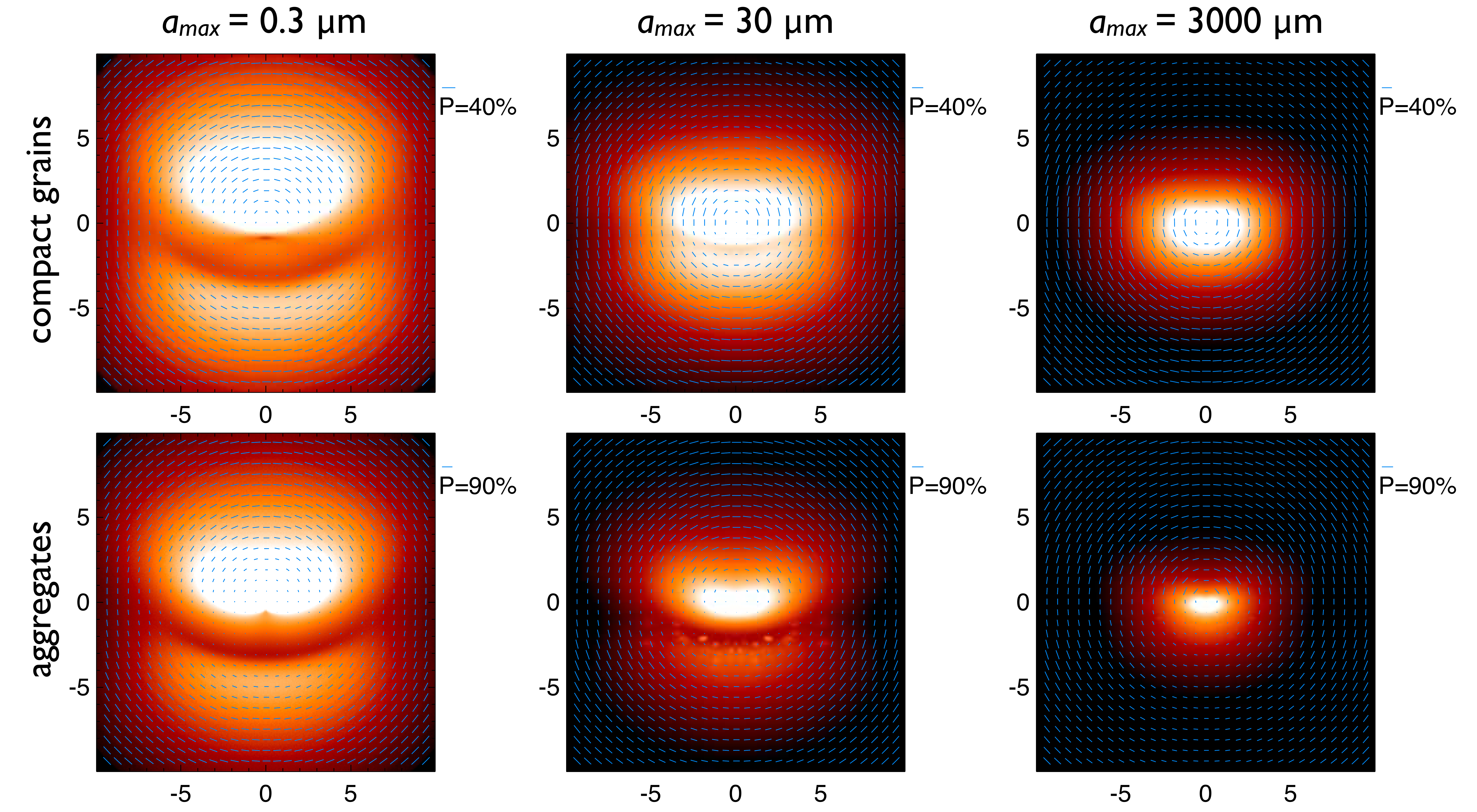}}}
\caption{Comparison of the polarized intensity maps for different grain sizes. All panels are for the well mixed, i.e. flaring, disk models. Upper panels are for the compact grains, lower panels for the fluffy aggregates. The maximum particle size in the size distribution increases from left to right. Spatial scale is in arcseconds. The length of the vectors is proportional to the degree of polarization. The inclination of the disk is 55$^\circ$ with respect to pole on. Color scale is the same as in Fig.~\ref{fig:grain types}.}
\label{fig:grain sizes group I}
\end{figure*}

\begin{figure*}[!htb]
\centerline{\resizebox{0.9\hsize}{!}{\includegraphics{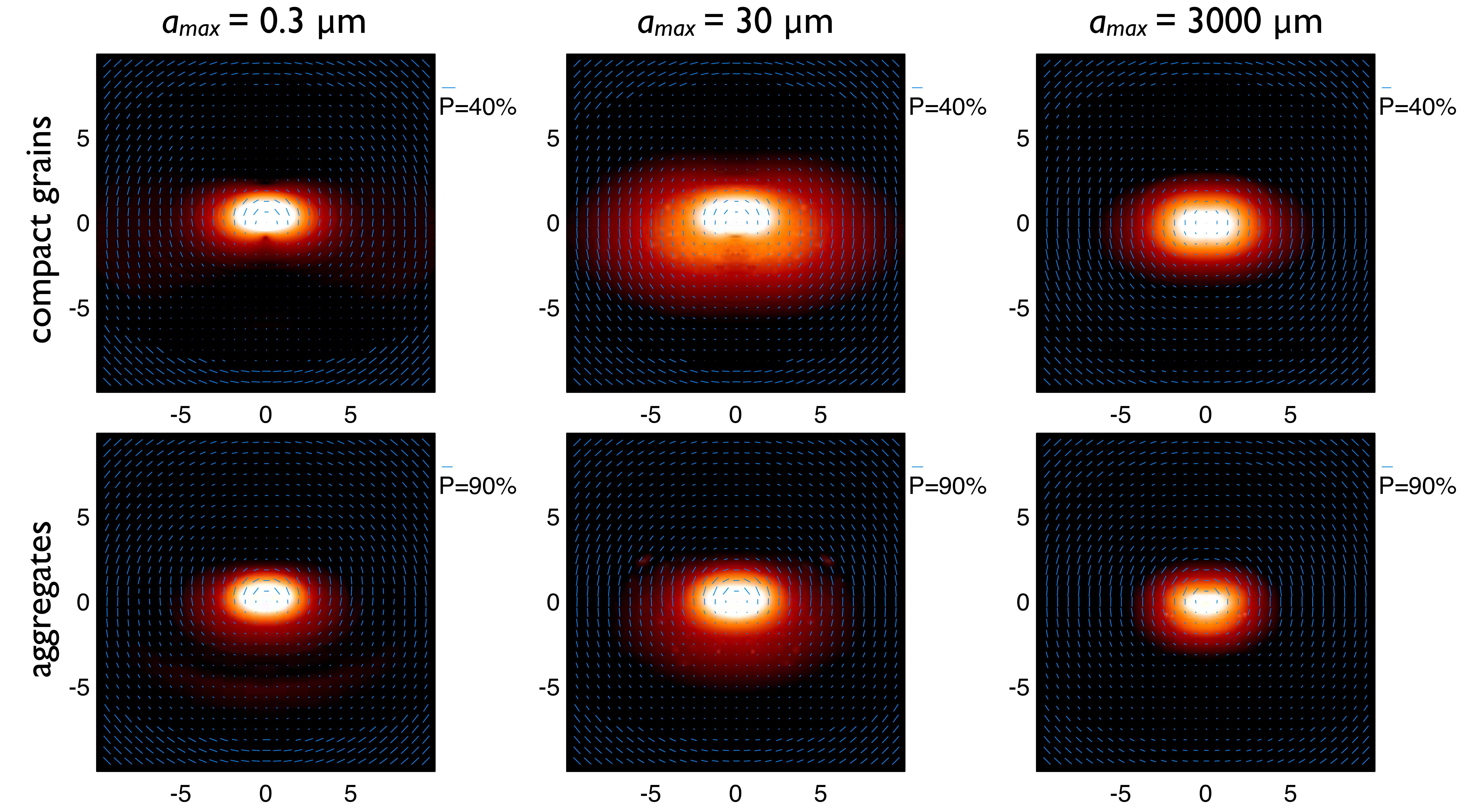}}}
\caption{Same as Fig.~\ref{fig:grain sizes group I}, but for the settled disk models.}
\label{fig:grain sizes group II}
\end{figure*}

\begin{figure*}[!htb]
\centerline{\resizebox{0.9\hsize}{!}{\includegraphics{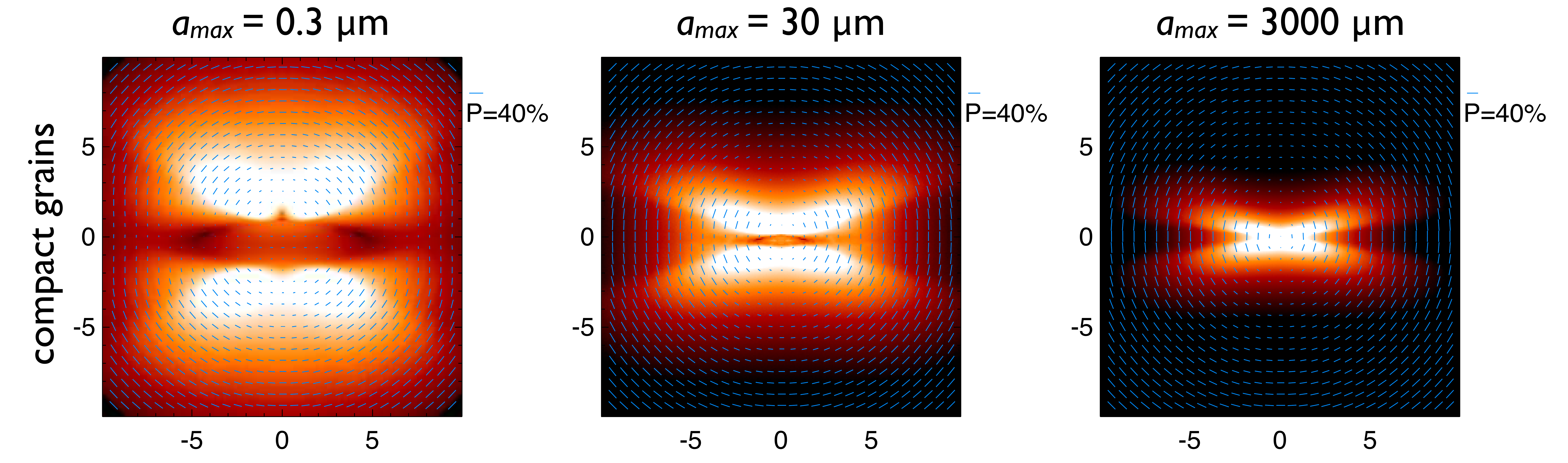}}}
\caption{Comparison of the polarized intensity maps for different grain sizes in the nearly edge on case. All panels are for the flared disk models with compact dust grains. The settled disks display only a very thin signature. The maximum particle size in the size distribution increases from left to right. Spatial scale is in arcseconds. The length of the vectors is proportional to the degree of polarization. The inclination of the disk is 85$^\circ$ with respect to pole on. Color scale is the same as in Fig.~\ref{fig:grain types}.}
\label{fig:edge on}
\end{figure*}

\section{Resulting full-resolution images}
\label{sec:results}

In this section we discuss the intrinsic polarimetric images of the disks and the effects of the different dust properties and disk parameters. All full-resolution model images are generated at a wavelength of $\lambda=600\,$nm. Though detailed studies of the color behavior of the polarized images can be a useful diagnostic, the general monochromatic behavior of the images is not very sensitive to the exact wavelength. We use these single wavelength images to represent the average broadband signal, so the broadband effects are fully taken into account for the simulation of the telescope and instrument, but the full-resolution model disk images are computed at a single wavelength.

In Fig.~\ref{fig:grain types} we present polarization maps for the different dust types. {We show here images for our two main grain types (compact grains and fluffy aggregates) and also for compact homogeneous spherical grains}. It is clear from this figure that the type of dust grain has a strong influence on the appearance of the disk. Especially in the upper part of the image, the homogeneous spherical dust grains display a behavior not seen for the more realistic dust models. {The homogeneous spherical grains display here a decreasing polarized intensity towards the upper central region of the disk. This feature is seen in a much milder form in the image of the fluffy aggregates, but generally the images for the other two grain shape models show smoother behavior in this part of the image.} Also, it is clear that the fluffy aggregates are fainter than the compact grain models. This is caused by the extremely forward-scattering nature of the fluffy aggregates, where only a small fraction of the light is scattered towards the observer. 

In Figs.~\ref{fig:grain sizes group I} and \ref{fig:grain sizes group II} we show the polarized intensity maps for the compact grains and the fluffy aggregates for different values of the maximum grain size in the size distribution. For the flared disk models (Fig.~\ref{fig:grain sizes group I}), it is clear that the brightness and spatial extent of the disk decreases when the particle size is increased. This is caused by the reduced opacity of the larger particles, which decreases the flaring angle of the disk. For the settled disk models we see a slightly different trend. Because the surface layer of the disk is here already in the shadow, the decrease in the height of the disk is not so important. More important in this case is the increase in the scattering albedo, causing the disk to become brighter and appear larger at first when the particle size is slightly increased. When the particle size is increased further, the decrease in the total opacity becomes dominant and the disk becomes fainter again.

\cite{2010A&A...518A..63M} {revived the discussion on} the `roundabout effect' for edge on disks in near infrared polarimetric imaging. In the dark lane of an edge on disk no first-order scattering can be observed since no starlight can reach these regions in the disk. Second-order scattering causes the polarization vectors not to be aligned perpendicular to the direction towards the central star, but parallel to the dark lane of the disk. {This effect was first noted by \citet{1988ApJ...326..334B}, after which \citet{1992ApJ...395..529W} argued that detecting this feature would be very hard because of the low surface brightness in this part of the image.} We also observe this effect in our optical polarization images of the most flaring edge-on disks. Figure~\ref{fig:edge on} shows the polarized intensity maps for the disks seen edge-on for different grain sizes. We only show the compact grain models, but the fluffy aggregate models show similar behavior. Clearly we see the roundabout effect in the flaring disk with only small grains. As soon as the grain size is increased, or the disk is settled, the effect disappears because the dark lane becomes too narrow. The region where the effect can be observed is very weak in polarized intensity as compared to the region above and below. This implies that, in observations with a limited spatial resolution, the effect will be washed away by PSF smearing.

\begin{figure*}[!htb]
\centerline{\resizebox{0.9\hsize}{!}{\includegraphics{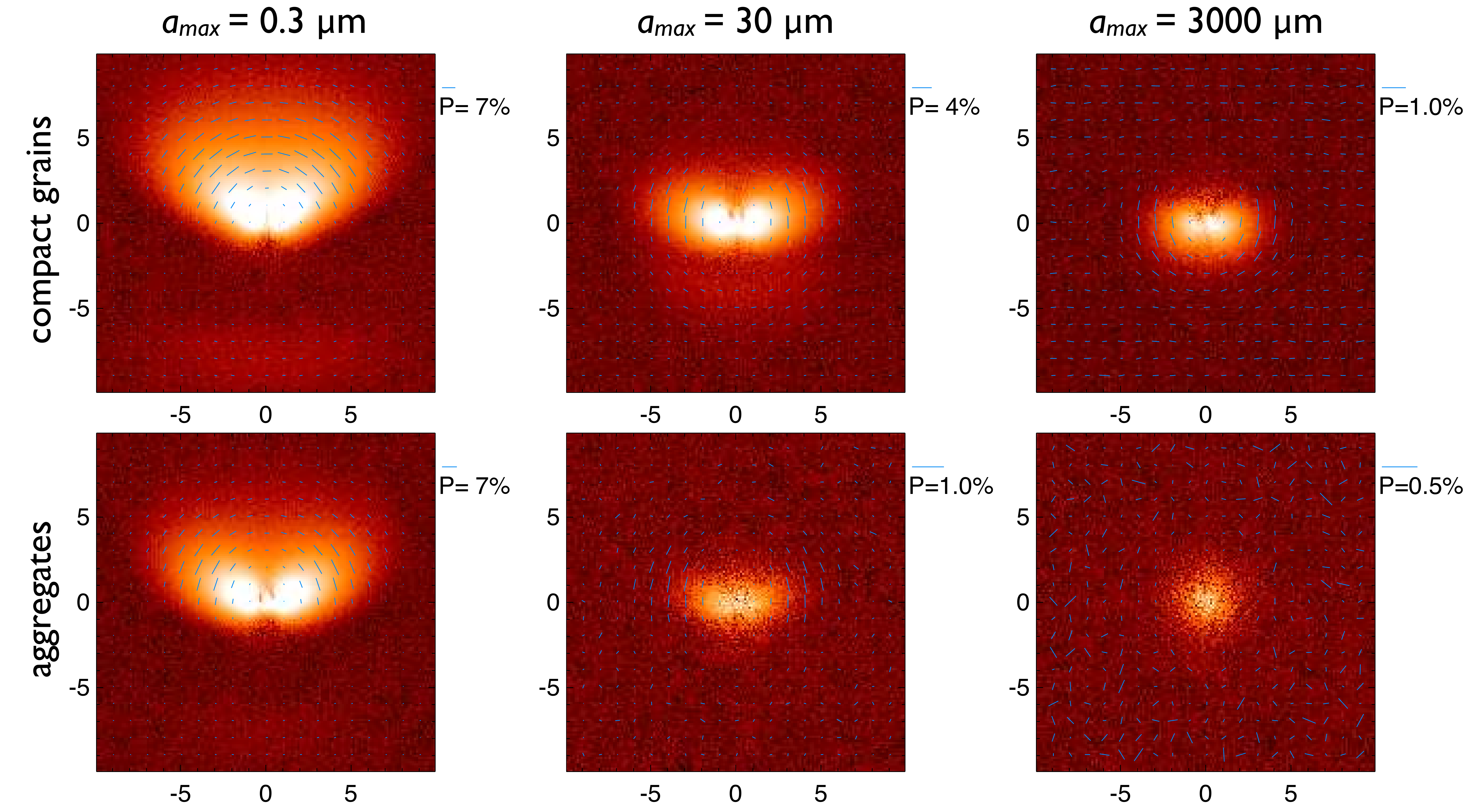}}}
\caption{Same as Fig.~\ref{fig:grain sizes group I}, but simulated as observed with ExPo at the WHT. Color scale is logarithmic, in arbitrary flux units, and the same in all images.}
\label{fig:exposim group I}
\end{figure*}

\begin{figure*}[!htb]
\centerline{\resizebox{0.9\hsize}{!}{\includegraphics{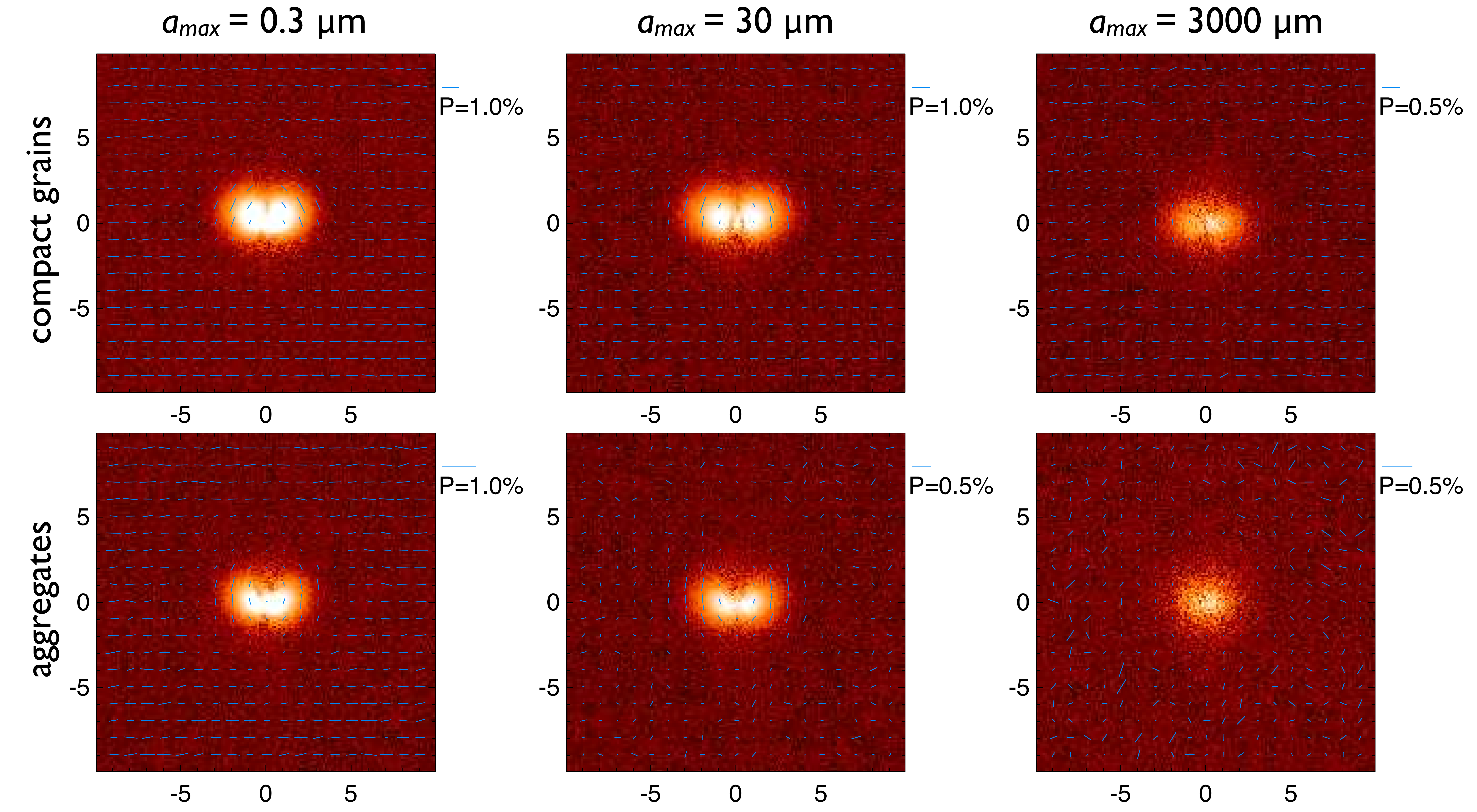}}}
\caption{Same as Fig.~\ref{fig:grain sizes group II}, but simulated as observed with ExPo at the WHT.}
\label{fig:exposim group II}
\end{figure*}

\section{Simulated observational images}
\label{sec:telescope effects}

We present images simulated for the ExPo instrument mounted at the 4.2 m William Herschel Telescope (WHT). The aim here is to get a feeling for the effects of atmospheric seeing, telescope diffraction, and noise on the interpretation of polarimetric images.

The final image observed through a telescope in polarization from a protoplanetary disk not only depends on the intrinsic degree of polarization caused by the grains but is also influenced by the depolarizing effects of the PSF of the central star. How strong this depolarizing effect is depends on the intensity of light scattered off the disk.
Below we discuss the effects of the different grain types on the intrinsic polarization image, as well as how these different models are affected by telescope effects such as diffraction, seeing, and noise.

In Figs.~\ref{fig:exposim group I} and \ref{fig:exposim group II} we show the different disk and dust types as they would be observed with ExPo at the WHT under good seeing conditions, i.e. a seeing at 500\,nm of 0.8''. There is a clear dependence of the detectibility on the particle size distribution and the vertical disk geometry. The larger and the more settled disks are much harder to detect, since they are intrinsically fainter.

The polarized intensity is the product of the total intensity and the degree of polarization. Due to the opening angle of the disk, the back side of the disk is closer to 90$^\circ$ scattering angle and thus has a higher degree of polarization (see also Fig.~\ref{fig:disk cartoon}). This can be seen in the high-resolution images by the longer vectors in the upper parts of the images. The front side of the disk is, however, much more forward scattering and thus has a higher total intensity. Because the central star is the dominant intensity source in the observational images, the degree of polarization now is basically the polarized intensity divided by the total intensity of the central star and not by the total intensity of scattered light. Looking at the observed settled disk models (i.e. Fig.~\ref{fig:exposim group II}), it can be seen that in the images for the compact grains, the degree of polarization still dominates (also see the length of the polarization vectors), and the back side of the disk is somewhat brighter in polarized intensity. Due to the highly forward-scattering nature of the fluffy aggregates, in that case the differences in total intensity dominate, and the front side of the disk is brighter in polarized intensity and in degree of polarization.

\begin{figure}[!tb]
\centerline{\resizebox{\hsize}{!}{\includegraphics{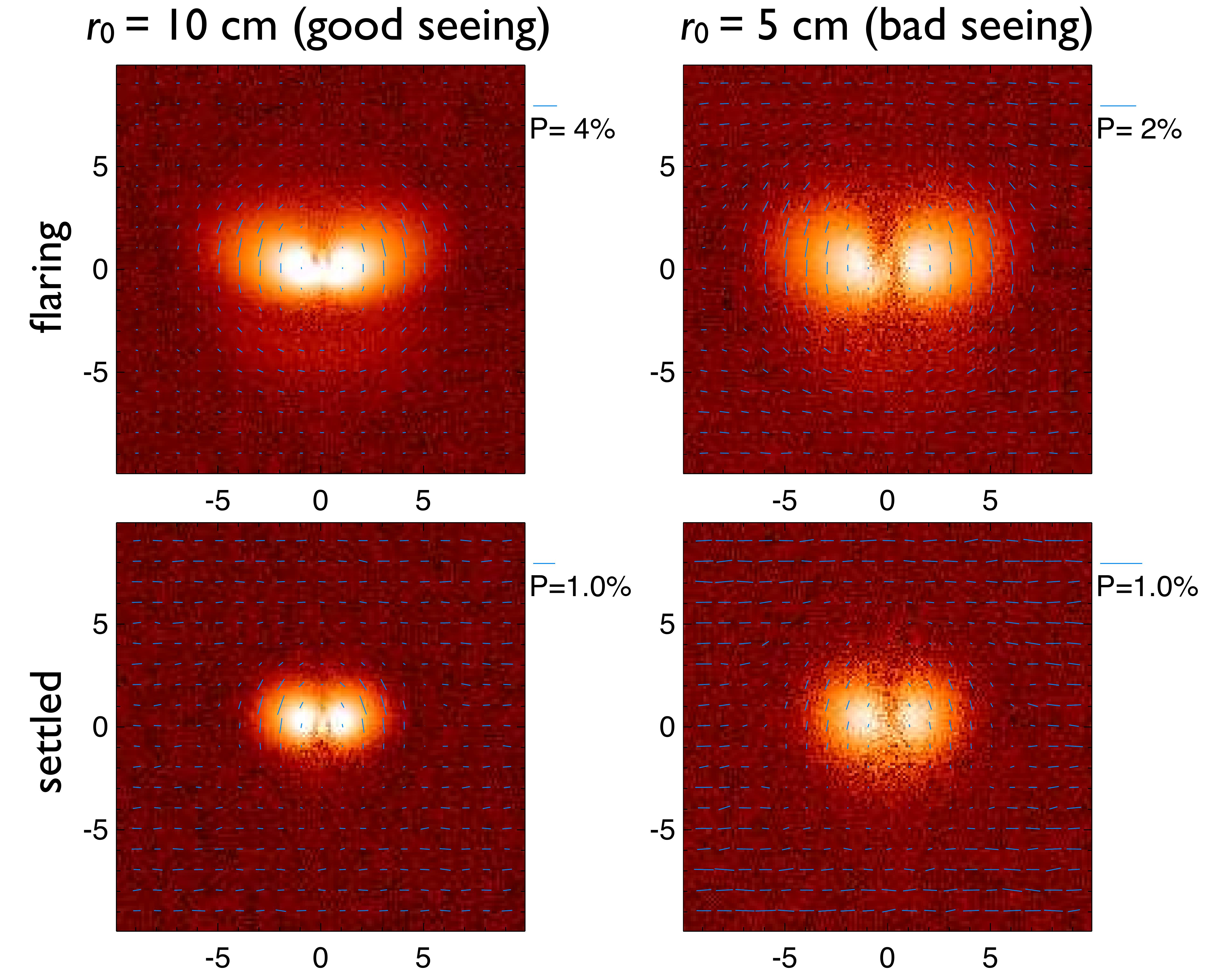}}}
\caption{Comparison of the polarized intensity maps as obtained under different seeing conditions. Upper panels are for a flared disk, lower panels for a settled disk. All images use $a_\mathrm{max}=30\,\mu$m and compact grains.}
\label{fig:seeing comparison}
\end{figure}

\begin{figure}[!tb]
\centerline{\resizebox{0.8\hsize}{!}{\includegraphics{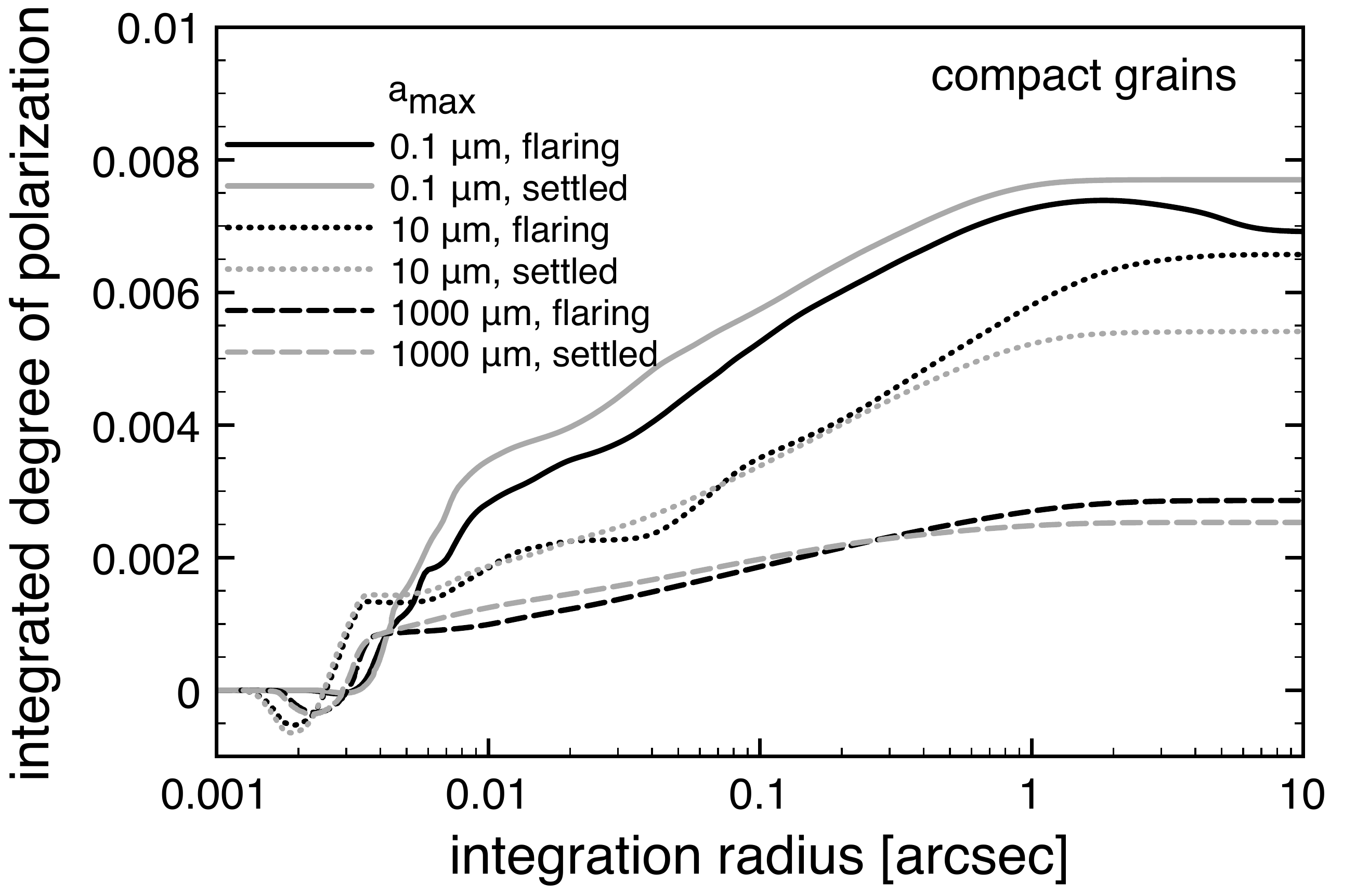}}}
\centerline{\resizebox{0.8\hsize}{!}{\includegraphics{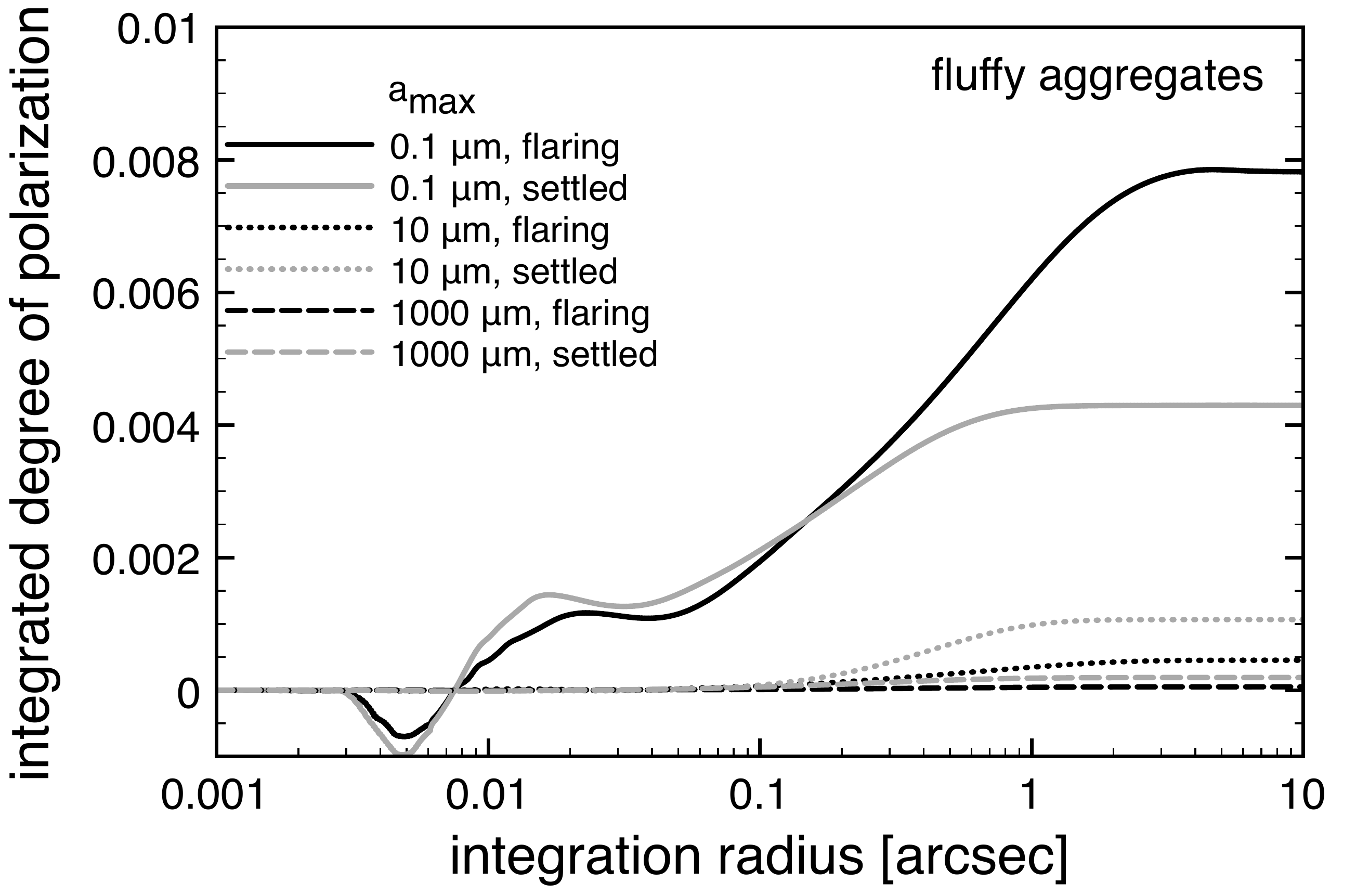}}}
\caption{The integrated degree of polarization as a function of the radius of integration. Upper panel is for the compact grains, lower panel for the fluffy aggregates. Black curves are for the flared disk models, grey curves for the settled disk models.}
\label{fig:inner pol}
\end{figure}

Since we run the full ExPo data analysis pipeline as discussed in \cite{2011A&A...531A.102C}, we also correct for any instrumental polarization. This is done
by assuming that the signal from the central resolution element, i.e. the central star, should be unpolarized. In the simulation of the instrument we do
not add polarization caused by the telescope, but we include instrumental polarization due to different transmission coefficients (see Appendix).
Nevertheless, the central resolution element is polarized because of scattering from the innermost regions of the disk. Since the disk is inclined,
scattering by this innermost region has a non-zero, integrated polarization (see also Fig.~\ref{fig:inner pol}). This polarization is then `corrected' under
the assumption that it is instrumental, and thus subtracted from the entire image. This causes the vertical depolarized region in the images where the
outer disk is weak. This also causes the well oriented polarization vectors in the noise regions of the images for the cases where the disk is very weak
(especially clear in Fig.~\ref{fig:exposim group II}). The amplitude of this effect strongly depends on the seeing conditions. In
Fig.~\ref{fig:seeing comparison} we show two disk models under different seeing conditions. It is clear that the appearance is different and the overcorrection
of the instrumental and inner disk polarization is larger in the case of worse seeing.

In Fig.~\ref{fig:inner pol} we show the degree of polarization of the inner resolution elements as a function of the size of the resolution element. We do this by integrating both Stokes I and Q out to a distance $R$ from the central star. The degree of polarization shown in Fig.~\ref{fig:inner pol} is the ratio of these two integrated values. This shows that increasing the spatial resolution also significantly reduces the effect of polarization from the central resolution element. For images limited by an atmospheric seeing of $\sim1''$, we see that the inner resolution element can be polarized by up to $\sim0.8$\%. By adding an extreme adaptive optics system increasing the spatial resolution to $0.02''$, this degree of polarization can be reduced significantly. In the case of fluffy aggregates that have grown to significant sizes, we see that the integrated degree of polarization is very low. This is because the total scattered intensity is dominated by forward scattering, which has a very low degree of polarization. Although the polarization of the central resolution element is low, the spatially resolved degree of polarization is also extremely low.
The large variations in the curves in Fig.~\ref{fig:inner pol} show that we do have a good diagnostic for the grain and disk properties when we have sufficient resolution and the images are properly calibrated.

\subsection{Detectability}

For a disk to be detectable with an imaging polarimeter at a 4m class telescope without adaptive optics, we need the disk to be extended enough to escape the PSF of the central star and bright enough to rise above the photon and readout noise. Treating the models as observations we have determined which disk and dust parameters allow detection with this instrument. For this we assumed a total integration time of about five minutes per polarization state. The visual magnitude of the star, as derived from the input parameters in Table~\ref{tab:parameters}, is $V=7$. {All disks are assumed to be at a distance of 150\,pc. The detectability of the disks is based on the signal-to-noise ratio (S/N) of the polarized intensity images. 
The orientation of the polarization vectors depends on this parameter, so the polarization vectors calculated for an image with poor S/N will be randomly oriented (see, for instance, Fig.~\ref{fig:exposim group I}, bottom righthand image). On the other hand, an image with high S/N will show a clear pattern on the orientation of its polarization vectors. For disks that are hardly detected, we also looked at the images for Stokes Q and U to see if a so-called `butterfly' pattern, typical of polarimetric images of circumstellar material, emerges above the noise.}

The results are summarized in Table~\ref{tab:detectability}. We see that disks that have undergone significant grain growth are hard to detect, especially when the turbulence in the disk is not strong enough to mix up the larger grains to the disk surface layers, i.e. in the settled disk models. Also, we notice that the seeing conditions play a major role in the detectability. By eliminating this factor, an adaptive optics system will greatly increase the detection capabilities of imaging polarimeters.

\begin{table}
\centering
\begin{tabular}{l c  c  c}
\multicolumn{1}{r}{$a_\mathrm{max}$}	&	$0.3\,\mu$m	& $30\,\mu$m	& $3000\,\mu$m \\
\hline
\multicolumn{1}{c}{Good seeing ($r_0=10$\,cm)} \\
\hline
Flared disk, compact grains	&	+						&	+							&	+	\\
Flared disk, fluffy aggregates	&	+						&	+/-							&	-	\\
Settled disk, compact grains	&	+						&	+							&	-	\\
Settled disk, fluffy aggregates	&	+						&	+							&	-	\\
\hline
\multicolumn{1}{c}{Bad seeing ($r_0=5$\,cm)} \\
\hline
Flared disk, compact grains	&	+						&	+							&	-	\\
Flared disk, fluffy aggregates	&	+						&	-							&	-	\\
Settled disk, compact grains	&	+						&	+/-							&	-	\\
Settled disk, fluffy aggregates	&	+						&	-							&	-	\\
\end{tabular}
\caption{Detectibility of the different disk and dust types with ExPo at the WHT. See Table~\ref{tab:parameters} for the details of the model.}
\label{tab:detectability}
\end{table}

\subsection{Disk gaps}

Many systems have been observed, directly or indirectly, to have depleted inner disks or cleared out gaps \citep[see e.g.][]{2003A&A...401..577B, 2007ApJ...664L.107B, 2008ApJ...682L.125E, 2010ApJ...718L..87T, 2010ApJ...718.1200M, 2011ApJ...732...42A, 2011A&A...528A..91V}. One of the possible causes for this is clearing by a planet in the disk, and thus direct detection of such disk gaps is very important for constraining planet formation scenarios and timescales. To see if we can detect such disk gaps and see what it would generally look like in a polarimetric imaging observation, we ran a disk model with a cleared out gap from 120 to 180\,AU and an inner disk with a depletion of a factor of ten for radii smaller than 120\,AU \citep[a parameterization very similar to the one adopted by][]{2011ApJ...732...42A}. We did not aim to span parameter space for this case, but rather to provide an example of inner disk clearing. {The radii of the gap were chosen to be close to what one would expect to be able to detect given the spatial resolution we had. Tests with closer in gaps show that, indeed, the radii presented here is at the edge of the detection capabilities given the simulated spatial resolution.}

\begin{figure}[!tb]
\centerline{\resizebox{\hsize}{!}{\includegraphics{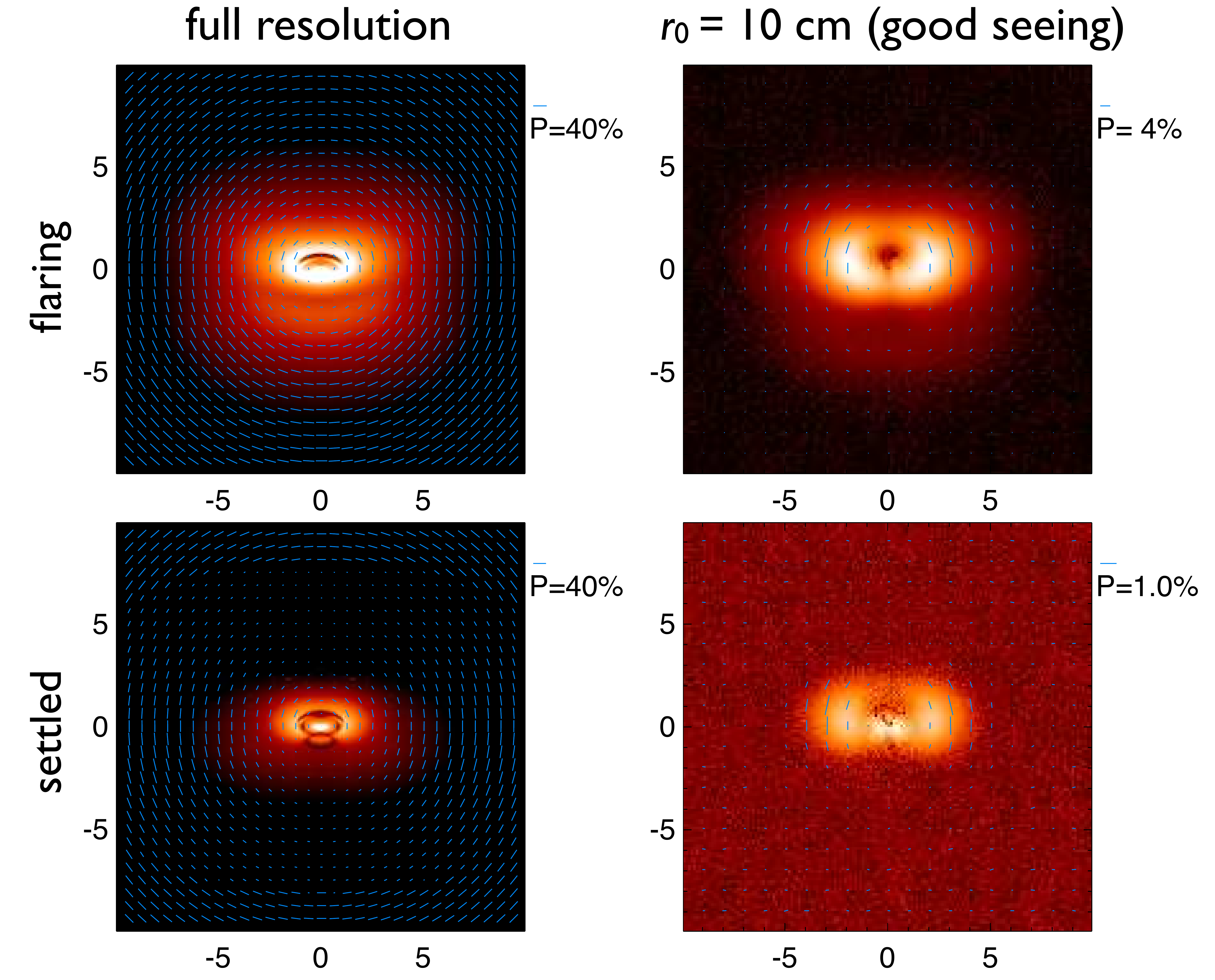}}}
\caption{Full resolution images (left panels) and simulated observations (right panels) of models with a depleted inner disk. The innermost regions (radii $<120\,$AU) are depleted by a factor of 10, while from $120$ to $180\,$AU the disk is cleared of dust. Upper panels are for a flared disk, lower panels for a settled disk. All images are using $a_\mathrm{max}=30\,\mu$m and compact grains. The color scale is altered with respect to the previous images to enhance details in the bright inner disk region.}
\label{fig:disk gaps}
\end{figure}

The images for compact grains with $a_\mathrm{max}=30\,\mu$m are shown in Fig.~\ref{fig:disk gaps}, where we show both the full-resolution
images (left) and the simulated observations (right). For the instrument simulations we increased the integration time by a factor of ten to suppress
the photon noise in the inner disk regions. As we can see, the depletion of the inner region is clearly detected. Comparing to the lefthand panels of
Fig.~\ref{fig:seeing comparison} (which shows the same models without depleted inner disks), we see a very clear difference. We do note that
the inner disk is undetectable for the flaring disk case in the simulated observations, while it is still clearly visible in the full-resolution
images. This is caused by the fact that the entire inner disk is inside the central resolution element (0.8'' seeing). As a consequence the polarization
signal caused by the inner disk is considered instrumental by the data reduction pipeline and is subtracted. We also computed images where
we placed the gap closer in, and we find that an inner radius of the outer disk of $\sim180\,$AU (or 1.2'') is close to the limit where we are still able to
clearly identify the disk gap in the image. For radii of the inner edge of the outer disk larger than $\sim100\,$AU, we do see significant changes to the
image, but they become increasingly more difficult to identify as a clear signal from a disk gap as the radius decreases.

\section{Discussion and conclusions}
\label{sec:conclusions}

The appearance of circumstellar disks in polarized light depends heavily on the properties of both the disk and the composing
dust particles. We modeled the observable signature of protoplanetary disks surrounding young, intermediate-mass stars with various
dust and disk properties. We focus on the effects of grain structure and size-dependent grain settling. We also studied the effects of atmospheric blurring and observational data processing.
We conclude the following.
\begin{itemize}
\item There is a significant difference between the appearance of a disk with only compact grains and a disk containing fluffy aggregated dust particles. 
\item The commonly employed homogeneous sphere model gives significantly different results from more realistic models, because the observed signal depends strongly on the properties of these more realistic models, thus providing a potentially powerful diagnostic. The proper use of this diagnostic is not always trivial.
\item Due to the extreme forward-scattering nature of fluffy aggregates, where only a small fraction of the radiation is scattered towards the observer, a disk containing these kinds of particles appears significantly dimmer in scattered light.
\item The diagnostic power of imaging polarimetry is a strong function of spatial resolution. Atmospheric seeing blurs many details needed for
a thorough analysis of dust properties, and thus mainly allows the global disk parameters to be deduced.
\item The `roundabout effect' for edge-on disks as discussed by \cite{2010A&A...518A..63M} for near-IR polarimetry was only found in our study for an extremely flaring disk containing only small grains. Since the effect appears in the very low-intensity regions of the image, it disappears when
we consider observations with limited spatial resolution.
\item The accuracy of the correction for instrumental polarization using the central star as calibration source strongly depends on the spatial resolution.
In the seeing-limited case, the central resolution element can have a degree of polarization of up to $\sim0.8$\%. This degree of polarization quickly
decreases when the spatial resolution of the observations is increased.
\end{itemize}


\begin{appendix}

\section{Instrument simulation}
\label{app:instrument}

Here we discuss the details of the instrument simulations. The instrument simulator was developed to simulate observations
with the Extreme Polarimeter (ExPo), which is a visitor instrument of the 4.2 m WHT.

\subsection{Wavefront generation}
\label{sec:seeing}

\begin{figure}
  \centering
     \resizebox{0.8\hsize}{!}{\includegraphics[trim =  60 310 40 40]{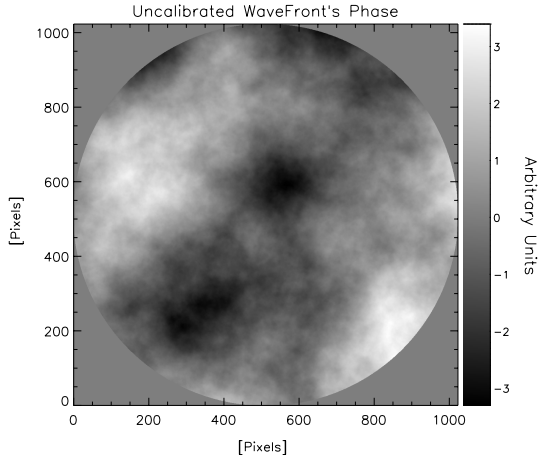}}

   \caption{Phase of a wavefront following the Kolmogorv model of turbulence.}
  \label{fig:uncal_wf}
\end{figure}

To generate wavefronts, we make use of a model of turbulence first proposed by \citet{Kolmogorov} and later developed by
\citet{Tatarski} and \citet{Fried}. According to this model, a flat wavefront traveling through the turbulent atmosphere will modify its
phase, while the change in its amplitude is negligible compared to the phase fluctuations. 
After it travels through the telescope, the amplitude of the wavefront is proportional to the telescope's aperture. The power spectrum of the phase fluctuations
of a wavefront traveling  through the atmosphere is given by
\begin{equation}
\Phi(k)= \frac{0.023}{r_{0}^{5/3}}k^{-11/3},
\label{eq:noll}
\end{equation}
where $r_{0}$ is \textit{Fried's Parameter} and $k$ the wave number. The parameter $r_{0}$ represents the circular aperture over which the wavefront
phase variance is equal to 1\,$\mathrm{rad}^{2}$ for the case of Kolmogorov turbulence. Therefore, the higher $r_{0}$, the more stable the atmosphere,
or, in other words, the better the \textit{seeing} is. For instance, a value of $r_{0}$ of about 10 cm produces 1 arcsec seeing at $\lambda = 500$ nm
\citep[i.e.][]{seeing_1}. The value of $r_{0}$ depends on the wavelength, following a power law:  $r_{0} \propto \lambda^{6/5}$. Figure~\ref{fig:uncal_wf}
shows an example of an \textit{uncalibrated} wavefront phase generated from Eq.~\ref{eq:noll}.

The timescale at which one of these wavefronts remains constant, the \textit{coherence time}, at optical wavelengths is only a few milliseconds \citep{La_Palma_Seeing_1994, Lucky_Thesis, Teide_Seeing_2009}. The single frame PSF has two main components:
\begin{itemize}
\item \textit{Tilt,} producing random motion of the whole image.
\item \textit{Roughness,} producing the observed speckle pattern.
\end{itemize}
When the exposure time is similar to the coherence time, the image 
motion due to the tilt component can be removed in the data processing by proper centering of each frame. \citet{Noll_76} showed that the variance  $\sigma^{2}$ of 
the wavefronts can be expressed in terms of the telescope diameter, D, and the Fried parameter, $r_0$
\begin{equation}
\sigma^{2} = 0.134 \left( \frac{D}{r_{0}} \right)^{5/6}.
\label{eq:noll2}
\end{equation}
The standard deviation $\sigma$ of the wavefront phase $\varphi$, expressed in wavelength units, can be obtained from the variance as $\sigma_{\lambda} = 2\pi\sigma$
\begin{equation}
\sigma_{\lambda} = \frac{0.134^{1/2} (D/r_{0})^{5/6}}{2\pi} = 0.0582 \left( \frac{D}{r_{0}}\right)^{5/6}.
\label{eq:noll3}
\end{equation}

In our simulation, D is the diameter of the WHT main mirror (4.2 meters), and three different values of $r_{0}$, (5, 7, and 10 cm) are used to account for bad, medium, and good seeing at a reference wavelength of 500 nm, $\lambda_{500}$. From Eq~\ref{eq:noll2} and the dependence of $r_{0}$ with wavelength, 
$\sigma_{\lambda}$ can be computed at different wavelengths from our reference as
\begin{equation}
\sigma_{\lambda} = \sigma_{\lambda_{500}} \cdot \frac{\lambda_{500}}{\lambda}.
\label{eq:sigma_wf}
\end{equation}

Wavefronts at different wavelengths are calibrated according to Eqs.~\ref{eq:noll2} and \ref{eq:sigma_wf}. A set of wavefronts at different wavelengths 
is then generated for each of the three different $r_{0}$ values used in this simulation.

\begin{table}
\label{tab:dd terms}
\centering
\begin{tabular}{ l  c  }
Parameter				&	Value \\
\hline\hline
Telescope diameter		&	4.2 [m]				\\
Wavelength range		&	450-700 [nm]		\\
Exposure time			&	0.028 [seconds]		\\
Pixel size			&	0.078 ["/pixel]		\\
Grid size			&	$2048\times2048$ [pixels] \\
\hline
\end{tabular}
\caption{Simulation parameters}
\label{tab:simterms}
\end{table}

\subsection{PSF generation}

\begin{figure}
  \centering
     \resizebox{\hsize}{!}{\includegraphics[trim = 100 400 105 140]{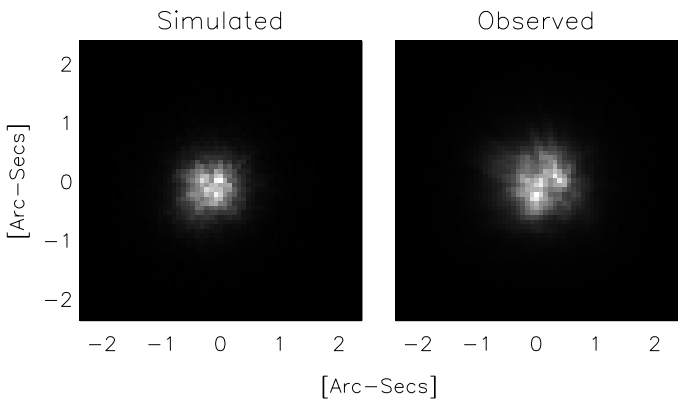}}

   \caption{Left: Simulated PSF with seeing $\approx 0.8"$, 0.028 seconds exposure time. Right: Observed PSF under the same conditions.}
  \label{fig:psdcomparison}
\end{figure}

To generate a PSF, we first simulate the aperture of the WHT, taking the spiders of the telescope and the central obscuration
of the main mirror into account. Once the telescope aperture and the wavefront are computed, a monochromatic PSF is calculated according to Eq.~\ref{eq:psf}.
A broadband PSF ($\mathrm{PSF}_{bb}$) is calculated as the sum of monochromatic PSFs calculated over the range of 400 nm to 700 nm, in steps of 10 nm:
\begin{equation}
\mathrm{PSF}_{bb} = \sum_{\lambda = 400,10}^{700} \left \arrowvert  \mathcal{F} \{  A \cdot e^{i\varphi(\lambda)}   \}   \right \arrowvert^{2}.
\label{eq:sigma_wf2}
\end{equation}
This PSF produces the speckle pattern obtained when observing at exposure times that are close to the coherence time (a few milliseconds). The PSF measured when observing
at longer exposure times will be the sum of the short-exposure broadband PSFs. To simulate ExPo-like PSFs, we set the coherence time to 9.3 milliseconds, and then generate
a 28 millisecond exposure PSF as the sum of three statistically independent short-exposure broadband PSFs:
\begin{equation}
\mathrm{PSF}_{\mathsc{ExPo}} = \sum_{i = 1}^{3} \mathrm{PSF}_{bb}^{i}.
\label{eq:sigma_wf3}
\end{equation}
These calculations are computed on a $2048\times2048$ pixels grid. The pixel size of this simulation is determined by the Nyquist frequency: $N_{\nu} = (\lambda/2D)\cdot206265$, which produces a pixel size of 0.0122 ["/pixel], for a 4.2 meter telescope and a central wavelength of 500 nm.
The original  $2048\times2048$ pixel grid is then binned to an smaller grid to produce simulated images at the ExPo pixel size (0.078 ["/pixel]).
Figure~\ref{fig:psdcomparison} shows an example of a PSF for both simulated (left) and real (right) data (0.8" seeing, 0.028 seconds exposure time).

\begin{figure}
  \centering
     \resizebox{\hsize}{!}{\includegraphics[trim = 100 430 145 175]{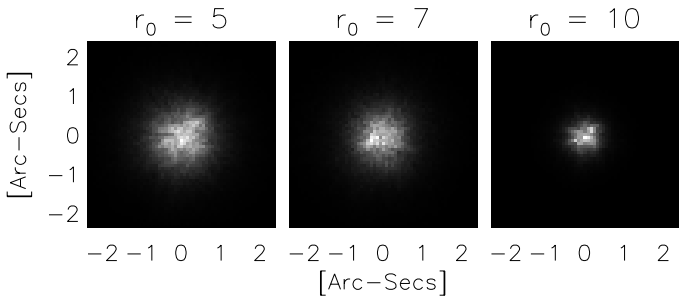}}

   \caption{Simulated speckle pattern for different values of $r_{0}$, given in centimeters.}
  \label{fig:speckle_comparison}
\end{figure}

A total of 100 different $\mathrm{PSF}_\mathsc{ExPo}$ are produced for each of the three different seeing conditions tested here. 
Figure~\ref{fig:speckle_comparison} shows an example of three different broadband PSF's calculated for bad 
(left), normal (center), and good (right) seeing.

\subsection{Dual beam simulation}

As a dual beam instrument, ExPo produces two simultaneous images with opposite polarization states that are imaged onto a CCD. We refer to these two images as \emph{left} and \emph{right} images. These measurements are modulated by a 
Ferro-electric liquid crystal (FLC), which switches between two orthogonal polarization states (A and B). Two
different frames, containing four different images are produced at the end of one FLC cycle:
\begin{equation}
\mathrm{A_{L}} = 0.5(I + P)_{0} *\mathrm{PSF_{A}},
\label{eq:a_l}
\end{equation}
\begin{equation}
\mathrm{A_{R}} = 0.5(I - P)_{0} * \mathrm{PSF_{A}},
\label{eq:a_r}
\end{equation}
\begin{equation}
\mathrm{B_{L}} = 0.5(I - P)_{0} * \mathrm{PSF_{B}},
\label{eq:b_l}
\end{equation}
\begin{equation}
\mathrm{B_{R}} = 0.5(I + P)_{0} * \mathrm{PSF_{B}},
\label{eq:b_r}
\end{equation}
where neither instrumental polarization nor instrumental effects are considered for the sake of simplicity, and  $I_{0}$ and $P_{0}$ represent the total intensity and polarization 
(Stokes $Q$ or Stokes $U$) of the observed target, as seeing without atmosphere and diffraction effects, respectively. $\mathrm{PSF_{A}}$ and $\mathrm{PSF_{B}}$ are the short exposure PSF 
for the A and B frames, respectively. These images are given in mJy units, and they are converted to \textit{counts} by considering the telescope area, exposure time,
filter transmission (Johnson V filter is used here), atmospheric + instrument absorption, and CCD efficiency. 

The \textsc{poidev} function from the NASA IDL library is used to simulate the photon noise for each of these images. Real readout noise from ExPo measurements 
is added to the simulated data. Instrumental polarization is simulated by including two transmission coefficients for each of the beams ($\mathrm{T_{L}},\mathrm{T_{R}}$)
and another two coefficients to account for the FLC transmission ($\mathrm{T_{A}},\mathrm{T_{B}}$). The value of these coefficients is listed in Table~\ref{tab:transmission}.
\begin{table}
\centering
\begin{tabular}{ c  c  }
Coefficient			&	Value \\
\hline\hline
$\mathrm{T_L}$		&	0.48			\\
$\mathrm{T_R}$		&	0.52			\\
$\mathrm{T_A}$		&	0.998		\\
$\mathrm{T_B}$		&	1.002		\\
\hline
\end{tabular}
\caption{Transmission coefficients measured from ExPo instrument.}
\label{tab:transmission}
\end{table}
To mimic guiding problems, a random shift with a maximum amplitude of ten pixels is applied to each of the simulated images. The final simulated images are then described by
\begin{equation}
\mathrm{A_{L}} = \mathrm{T_{A} \cdot T_{L}} \cdot(I + P)_{0} *\mathrm{PSF_{A}}\cdot M(x,y) + Ph_{AL}+ RO_{AL},
\label{eq:a_l2}
\end{equation}
\begin{equation}
\mathrm{A_{R}} = \mathrm{T_{A} \cdot T_{R}} \cdot(I - P)_{0} *\mathrm{PSF_{A}}\cdot M(x,y) + Ph_{AR}+ RO_{AR},
\label{eq:a_r2}
\end{equation}
\begin{equation}
\mathrm{B_{L}} = \mathrm{T_{B} \cdot T_{L}} \cdot(I - P)_{0} *\mathrm{PSF_{B}}\cdot M(x,y) + Ph_{BL}+ RO_{BL},
\label{eq:b_l2}
\end{equation}
\begin{equation}
\mathrm{B_{R}} = \mathrm{T_{B} \cdot T_{R}} \cdot(I + P)_{0} *\mathrm{PSF_{B}}\cdot M(x,y) + Ph_{BR}+ RO_{BR},
\label{eq:b_r2}
\end{equation}
where $M(x,y)$ represents the image-shifting function, $Ph$ is the photon noise, and $RO$ represents the readout noise.

We finally run the full data reduction pipeline for the simulated observations to obtain the final images presented in Figs.~\ref{fig:exposim group I}, \ref{fig:exposim group II}, and \ref{fig:seeing comparison}

\end{appendix}

\bibliographystyle{aa}

\end{document}